\documentclass[preprintnumbers, aps, floatfix, onecolumn, preprintnumbers, letterpaper, superscriptaddress,nofootinbib,11pt]{revtex4}
\usepackage{dcolumn}
\usepackage{graphicx}
\usepackage{amsmath}
\usepackage{amsfonts}
\usepackage{amssymb}
\usepackage{microtype}
\usepackage{subfigure}
\usepackage{makeidx}
\usepackage{bm}
\usepackage{epsf}
\usepackage[colorlinks=true,citecolor=blue,linkcolor=black,urlcolor=black]{hyperref}
\usepackage{color}
\usepackage{multirow,dcolumn}
\usepackage{graphicx}
\usepackage{mathrsfs}
\graphicspath{{Images/}}
\def\PR#1{{Phys.\ Rev.\ D \bf #1}}
\def\PRL#1{{Phys.\ Rev.\ Lett.\ \bf #1}}

\begin{document}

\title{General Black Hole Solutions in (2+1)-dimensions with a Scalar Field  Non-Minimally  Coupled to Gravity}

\author{Zi-Yu Tang}
\email{tangziyu@sjtu.edu.cn}
\affiliation{Center for Astronomy and Astrophysics, School of Physics and Astronomy,
Shanghai Jiao Tong University, Shanghai 200240, China}
\affiliation{Collaborative Innovation Center of IFSA (CICIFSA),
Shanghai Jiao Tong University, Shanghai 200240, China}

\author{Yen Chin Ong}
\email{ycong@yzu.edu.cn}
\affiliation{Center for Gravitation and Cosmology, College of Physical Science
and Technology, Yangzhou University, Yangzhou 225009, China}

\author{Bin Wang}
\email{wang\_b@sjtu.edu.cn}
\affiliation{School of Aeronautics and Astronautics, Shanghai Jiao Tong University, Shanghai 200240, China}
\affiliation{Collaborative Innovation Center of IFSA (CICIFSA),
Shanghai Jiao Tong University, Shanghai 200240, China}
\affiliation{Center for Gravitation and Cosmology, College of Physical Science
and Technology, Yangzhou University, Yangzhou 225009, China}

\author{Eleftherios Papantonopoulos}
\email{lpapa@central.ntua.gr} \affiliation{Physics Division,
National Technical University of Athens, 15780 Zografou Campus,
Athens, Greece.}

\begin{abstract}
We discuss black hole solutions in (2+1)-dimensions  with a scalar field non-minimally coupled to Einstein's gravity in the presence of a cosmological constant and a self-interacting scalar potential. Without specifying the form of the potential, we find a general solution of the field equations, which includes all the known asymptotically anti-de Sitter (AdS) black hole solutions in (2+1)-dimensions as special cases once values of the coupling constants are chosen appropriately. In addition, we obtain numerically new black hole solutions and for some specific choices of the coupling constants we derive new exact AdS black hole solutions. We also discuss the possibility of obtaining asymptotically de Sitter black hole solutions with or without an electromagnetic field.
\end{abstract}

\maketitle


\section{Introduction}

Despite the wide application of the AdS/CFT correspondence (and more generally the gauge/gravity duality) in a variety of settings, such as quark gluon plasma and condensed matter systems, the properties of asymptotically AdS spacetimes are still far from being completely understood, even at the classical level. A good example of our lack of understanding is the stability issue of asymptotically AdS spacetimes. It came as a surprise that -- unlike Minkowski spacetime -- AdS spacetime is generically \emph{unstable} under small perturbations. For Minkowski spacetime, small perturbations eventually disperse to infinity and thus the spacetime is stable  \cite{ck}. In the case of AdS spacetime, however, due to the presence of timelike boundary at infinity, perturbations can bounce off and return to the bulk (recall that null rays can bounce back in a finite time), which in turn can focus enough energy to cause black hole formation.
The stability of $D$-dimensional AdS spacetime, here denoted $\text{AdS}_D$, for $D\geq 4$ was studied in \cite{br,jrb}. It was found that for arbitrarily small
perturbations of a massless scalar field minimally coupled to AdS gravity,  the  $\text{AdS}_D$ spacetime is unstable and finally results in black hole formation. To be more specific, this instability was understood to
be  the result of resonant transfer of energy from low to high frequencies \cite{dhs}. In (2+1)-dimensions, however, AdS gravity has a different behavior. It was found \cite{Bizon:2013xha} that for a large class of perturbations,  a turbulent cascade of energy to high frequencies still occur, which entails instability of  $\text{AdS}_3$, but cannot be terminated by  black hole formation because small perturbations have energy below the black hole threshold (due to mass gap of black hole).

As is well known, in (2+1)-dimensions the local geometry without the presence of any matter field remains trivial even if a cosmological term
is introduced, since the Einstein space is  a space of constant curvature in (2+1)-dimensions. Surprisingly, the presence of the cosmological constant introduces a scale which allows one to find a black hole solution in the absence of any scalar field. This is known as the BTZ black hole \cite{Banados:1992wn}. It is obtained by identifying
certain points of  the anti-de Sitter spacetime. The BTZ black hole is characterized by the mass,  angular momentum and
a negative cosmological constant, and has almost all the features of the Kerr-AdS black hole
in four-dimensional Einstein gravity.

In the context of gauge/gravity duality,  one could construct black hole solutions in the gravity sector that acquires hair below a critical temperature.  Even a static and spherically symmetric black hole with scalar hair requires a deep understanding of the  matter field near the  black hole horizon, as it should satisfy some physical requirements, e.g. being regular on the horizon, and decays
sufficiently fast towards infinity, so that the black hole shows its presence as (primary or secondary) hair in the boundary field theory with finite temperature.

The early attempts to couple a scalar field to gravity was first carried out in
asymptotically flat spacetimes. Black hole solutions were found \cite{BBMB} but the scalar field was divergent on the horizon and
they were unstable under scalar perturbations \cite{bronnikov}. {Introducing a cosmological constant we expect that the  scalar field  to have regular behavior  on the resulting black hole horizon, while all
possible  divergences to be hidden behind the horizon.} Hairy black hole solutions were found in asymptotically de Sitter spacetime with either a minimally or a non-minimally coupled
scalar field but they were unstable \cite{Zloshchastiev:2004ny,Torii:1998ir,Martinez:2002ru,Harper:2003wt,Dotti:2007cp}.
In the case of AdS spacetime with a
negative cosmological constant, stable solutions were found
numerically for spherical geometries \cite{Torii:2001pg,
Winstanley:2002jt,Gonzalez:2013aca} and an exact solution in asymptotically AdS
spacetime with hyperbolic geometry was presented in
\cite{Martinez:2004nb} and generalized later to include charge
\cite{Martinez:2005di}. In all the above solutions the scalar
field is conformally coupled to gravity. A generalization to
non-conformal solutions was discussed in \cite{Kolyvaris:2009pc}. Besides, three dimensional black holes in dilatonic gravity were also studied in \cite{Chan:1994qa,Dehghani:2017zkm,Hendi:2017mgb,Hendi:2017oka}.

A scale can also be introduced in the scalar sector of the theory. This can be done if in the
Einstein-Hilbert action there is a coupling of a scalar field to the
Einstein tensor. The derivative coupling has the dimension of
length squared and it was shown to act as an effective
cosmological constant \cite{Amendola:1993uh,Sushkov:2009hk}. Spherically symmetric black
hole solutions which are asymptotically anti-de Sitter were found \cite{Rinaldi:2012vy,Kolyvaris:2011fk,Kolyvaris:2013zfa,Babichev:2013cya,Charmousis:2014zaa,Charmousis:2015txa,Cisterna:2014nua}, thus evading the no-hair theorem for Horndeski theory \cite{Hui:2012qt}.

In (2+1)-dimensions black holes with scalar field non-minimally coupled to gravity are known to exist \cite{Martinez:1996gn, Henneaux:2002wm}.
Furthermore, both static \cite{Correa:2011dt} and rotating \cite{Natsuume:1999at,Correa:2012rc,Liu:2014eqa} hairy (2+1)-dimensional black holes were constructed by specifying some suitable forms of the potential. An exact dynamical and inhomogeneous solution that represents gravitational collapse was presented in \cite{Xu:2014xqa}. If an electromagnetic field is introduced then it is possible to find hairy charged black hole solutions in (2+1)-dimensions \cite{Xu:2013nia}. More recently exact dynamical and inhomogeneous solutions in (2+1)-dimensional AdS gravity with a conformally coupled scalar field was discussed in  \cite{Aviles:2018vnf}.

Gravity in (2+1)-dimensions is interesting in many ways. One of the principal advantages of working
in (2+1)-dimensions is that for simple enough topologies, this space can be characterized
completely and explicitly, which can help to  gain important insights into black hole physics and the
structure of quantum gravity.  Hairy black hole solutions in the presence of a cosmological constant and conformally coupled matter are good theoretical laboratories to examine further the gauge/gravity correspondence \cite{Maldacena:1997re} by connecting the three dimensional gravity bulk with the two dimensional boundary field theory and holographically relating to, e.g., condensed matter systems of two dimensional materials and superfluids.

In this work we study general relativity in (2+1)-dimensions with a scalar field non-minimally coupled to the gravity sector, with a general potential in the presence of a cosmological constant.  Our aim is -- \emph{without specifying the form of the potential a priori} -- to find a general solution of the field equations in the coupled Einstein-scalar field system, which can give a general hairy black hole solution. To this end, we solve the Einstein-scalar field equations with a static and spherically symmetric metric ansatz. The general solution we found can  be reduced to the known black hole solutions  \cite{Banados:1992wn,Martinez:1996gn,AyonBeato:2001sb,Xu:2013nia,Xu:2014xqa} depending on the value of the coupling constant. Going beyond the known solutions in the literature, we can also derive other new solutions. We also investigate the possibility of finding asymptotically dS black hole solutions in (2+1)-dimensions and we show that no such black hole exist, consistent with previous no-go results. However, surprisingly, we note that \emph{in the presence of an electromagnetic field} the solution found in \cite{Xu:2013nia} can give an asymptotically de Sitter black hole solution for a specific choice of parameters.

This work is organized as follows: in  Section II we discuss the general solution of a scalar field  non-minimally coupled to gravity, in Section III we show some explicit -- exact solutions -- of  asymptotically AdS black holes, in Section IV we discuss the possibility of finding dS black hole solutions and finally in the last Section we present our conclusions.

\section{General solution with a scalar field  non-minimally coupled to Einstein's gravity}

In this section we consider a scalar field non-minimally coupled to (2+1)-dimensional Einstein's gravity with a self-interacting scalar potential. The  action is
\begin{equation}
I=\int d^3 x \sqrt{-g}\left[\frac{R}{2}-\frac{1}{2}g^{\mu\nu} \nabla_\mu\Phi \nabla_\nu\Phi-\frac{1}{2}\xi R \Phi^2-U\left(\Phi\right) \right],\label{action}
\end{equation}
where $\xi$ is a coupling constant and $U\left(\Phi\right)$ is the self-interacting potential of the scalar field in which  the cosmological constant is included via $U\left(\Phi=0\right)=\Lambda$ (see details below).
The Einstein equations and the Klein-Gordon equation read, respectively,
\begin{eqnarray}
\left(1-\xi \Phi^2\right)G_{\mu\nu}&=&\nabla_\mu\Phi \nabla_\nu\Phi-g_{\mu\nu}\left(\frac{1}{2}g^{\alpha\beta} \nabla_\alpha\Phi \nabla_\beta\Phi+U\left(\Phi\right)\right)+\xi \left(g_{\mu\nu}\square-\nabla_\mu\nabla_\nu\right)\Phi^2, \\
\square \Phi&=& \xi R \Phi+\frac{dU(\Phi)}{d\Phi}~, \label{KG}
\end{eqnarray}
where $G_{\mu\nu}$ denotes the Einstein's tensor.
We consider a spherically symmetric  ansatz for the metric
\begin{equation}
\text{d}s^2=-f(r)\text{d}t^2+\frac{\text{d}r^2}{f(r)}+r^2 \text{d}\theta^2~. \label{ansatz}
\end{equation}
Then, the $t$-$t$ and $r$-$r$ components of the Einstein equations give
\begin{equation}
\Phi(r)=\left(\frac{A}{r+B}\right)^{\frac{\delta}{2}},\label{Phi}
\end{equation}
where $\delta\equiv 4\xi/\left(1-4\xi\right)$, with $A$ and $B$ being some constants.
We first consider $A>0$, $B>0$ and $\delta>0$, corresponding to $0<\xi<{1}/{4}$. Note that $\delta\neq -1$, since $\delta=-1$ would correspond to unphysical value of $\xi = \infty$.

The $r$-$r$ component of the Einstein equations then gives the metric function
\begin{equation}
f(r)=\frac{(r+B)^{1+\delta}\left[C_0-8(1+\delta)\int U(r)r \text{d}r \right]}{A^\delta\delta\left[\delta r-(r+B)\right]+4(1+\delta)(r+B)^{1+\delta}}~,
\label{Metric}
\end{equation}
where $C_0$ is an integration constant. (In our notation, the integration constant is written explicitly, i.e., the integral $\int U(r)r \text{d}r $ itself is understood to be \emph{without} a constant term.)

If we substitute the scalar field (\ref{Phi}) and the metric function (\ref{Metric}) into the $\theta$-$\theta$ component of the Einstein equations, we would obtain an integral equation of the potential $U(r)$:
\begin{equation}
C_0-8(1+\delta)\int U(r)r \text{d}r=a(r)U(r)+b(r)U'(r)~,\label{eqIU}
\end{equation}
where
\begin{eqnarray}
a(r)&=&c(r) \left[A^\delta \delta  (3 B-\delta  r+r)-4 (\delta +1) (B+r)^{\delta } (3 B-2 \delta  r+r)\right],\\
b(r)&=& c(r) (B+r)A^{-\delta}\delta^{-2}\left[A^{2\delta} \delta ^2 (B-\delta  r+r)+4 A^\delta \delta  (\delta +1) (\delta  r-2 (B+r)) (B+r)^{\delta }\right. \notag \\
&&\left. +16 (\delta +1)^2 (B+r)^{2 \delta +1}\right],\\
c(r)&=&-4 (\delta +1) r (B+r) \left[4 (\delta +1) (B+r)^{\delta +1}-A^\delta \delta  (B-\delta  r+r)\right]/\left\{A^{2\delta} B (\delta -1) \delta ^2 (B+r) \right. \notag \\
&& -2 A^\delta \delta  (\delta +1)\left[-4 B^2+B \left(-3 \delta ^2+\delta -4\right) r+(\delta -1)^2 \delta  r^2\right] (B+r)^{\delta }\notag\\
&& \left. +8 (\delta +1)^3 (\delta  r-2 B) (B+r)^{2 \delta +1}\right\}.
\end{eqnarray}

By equation (\ref{eqIU}), the metric function (\ref{Metric}) becomes
\begin{equation}
f(r)=\frac{(r+B)^{1+\delta}\left[a(r)U(r)+b(r)U'(r)\right]}{A^\delta \delta\left[\delta r-(r+B)\right]+4(1+\delta)(r+B)^{1+\delta}}~,\label{Metric2}
\end{equation}
which can be reduced to the solutions in \cite{Martinez:1996gn} and \cite{AyonBeato:2001sb} when $U(r)=\Lambda=-\ell^{-2}$ and $\delta=1$. Moreover, when $\delta=0$, we get the BTZ black hole solution \cite{Banados:1992wn}.

Taking the  derivative of Eq.(\ref{eqIU}) we obtain, with prime denoting derivative with respect to $r$, the second order differential equation
\begin{equation}
U''(r)+P(r)U'(r)+Q(r)U(r)=0~,  \label{simple}
\end{equation}
where
\begin{equation}
P(r)=\frac{a(r)+b'(r)}{b(r)}~,\,\,\,\,
Q(r)=\frac{a'(r)+8(\delta+1)r}{b(r)}~.
\end{equation}

A particular solution of Eq.(\ref{simple}) is
\begin{equation}
U_1(r)=\frac{8(\delta+1)(r+B)^{2+\delta}-\delta(\delta-1)(\delta-2)A^\delta r^2+4\delta(\delta-1)A^\delta B r-2\delta A^\delta B^2 }{8(\delta+1)(r+B)^{2+\delta}}~.
\end{equation}
Using Eq.(\ref{Phi}) for $\Phi$ and introducing the ratio  $\gamma\equiv B/A$, we have
\begin{eqnarray}
U_1(\Phi)=\frac{\delta  \left[\Phi ^2 \left(-\delta  \left(\gamma  \Phi ^{2/\delta }-1\right) \left(\gamma  (\delta +1) \Phi ^{2/\delta }-\delta +3\right)-2\right)+8\right]+8}{8 (\delta +1)}~. \label{u1}
\end{eqnarray}
In the following we will discuss various solutions for the potential $U(r)$.

\subsection{ AdS spacetimes with a stealth structure}

If we choose $U(r)=C_1 U_1(r)$ as the potential, then the cosmological constant is simply
\begin{equation}
\Lambda=U(\Phi=0)=C_1~,
\end{equation}
and the metric function becomes
\begin{equation}
f(r)=\frac{(r+B)^{1+\delta}\left[a(r)C_1 U_1(r)+b(r)C_1 U_1'(r)\right]}{A^\delta \delta\left[\delta r-(r+B)\right]+4(1+\delta)(r+B)^{1+\delta}}=-r^2 C_1=-r^2 \Lambda~,
\end{equation}
with constant curvature $R(r) \equiv 6C_1=6\Lambda$. It describes AdS spacetimes when $\Lambda=-\ell^{-2}<0$, where $\ell$ is the length of the AdS space. In fact, this solution has what is known in the literature as a ``stealth structure'' -- this kind of matter configurations have no influence on the geometry.
Black holes with stealth scalar field in (2+1)-dimensions have been studied in \cite{AyonBeato:2004ig}, their potential, which is obtained from the vanishing of energy-momentum, is the same as our solutions.

\subsection{ General AdS black hole solution }

Using Liouville's formula, we can obtain the other linearly independent particular solution to Eq.(\ref{simple}):
\begin{equation}
U_2(r)=U_1(r)\int U_1(r)^{-2}e^{-\int P(r)\text{d}r}\text{d}r~. \label{U2}
\end{equation}
We define
\begin{eqnarray}
\Bbb{P}(r):=e^{-\int P(r)\text{d}r}&=&\frac{B (1-\delta)}{2 r (B+r)^{\delta +3}}+\frac{A^\delta \delta ^3 (\delta +1) [(\delta -3) r-4 B]}{(B+r)^4 \left[A^\delta \delta -4 (\delta +1) (B+r)^{\delta }\right]^2} \notag \\
&&-\frac{\delta  (\delta +1) (-4 B+\delta  r+r)}{r (B+r)^3 \left[4 (\delta +1) (B+r)^{\delta }-A^\delta \delta \right]},
\end{eqnarray}
so that
\begin{equation}
U_2(r)=U_1(r)\int U_1(r)^{-2}\Bbb{P}(r)\text{d}r~.
\end{equation}
The general solution of Eq.(\ref{simple}) is
\begin{equation}
U(r)=C_1 U_1(r)+C_2U_2(r)=U_1(r)\left[C_1+C_2\int U_1(r)^{-2}\Bbb{P}(r)\text{d}r\right]~,
\end{equation}
where $C_1$ and $C_2$ are  constants. The  constant of integration coming from $\Bbb{P}(r)$ can be absorbed into $C_2$.

The potential has been determined from the Einstein equations without using the Klein-Gordon equation (\ref{KG}). Considering the Bianchi identity $\nabla_\nu G_\mu^\nu=0$ the Einstein equations  are equivalent to the on-shell condition $\nabla_\nu T_\mu^\nu=0$, which ensures that our potential is valid under the on-shell condition. We have also checked that the Klein-Gordon equation  is always satisfied.

The potential can be written as a function of the scalar field $\Phi$
\begin{eqnarray}
U(\Phi)=U_1(\Phi)\left[C_1+C_2\int U_1(\Phi)^{-2}\Bbb{P}(\Phi)\frac{2}{A \delta}\Phi^{1/\delta-1}\text{d}\Phi \right],
\end{eqnarray}
where
\begin{eqnarray}
\Bbb{P}(\Phi)&=&\frac{\gamma  (\delta -1) A^{-\delta -3} \Phi ^{\frac{8}{\delta }+2}}{2 \left(\gamma  \Phi ^{2/\delta }-1\right)}+\frac{\delta  (\delta +1) \Phi ^{6/\delta } \left(5 \gamma  \Phi ^{2/\delta }+\delta  \left(\gamma  \Phi ^{2/\delta }-1\right)-1\right)}{A^3 \left(A^2 \delta -\frac{4 (\delta +1) A^{\delta }}{\Phi ^2}\right) \left(\gamma  \Phi ^{2/\delta }-1\right)}\notag \\
&&-\frac{\delta ^3 (\delta +1) \Phi ^{6/\delta } \left(\gamma  \Phi ^{2/\delta }+\delta  \left(\gamma  \Phi ^{2/\delta }-1\right)+3\right)}{A \left(A^2 \delta -\frac{4 (\delta +1) A^{\delta }}{\Phi ^2}\right)^2}.
\end{eqnarray}

In the above expression an integral remains  but all functions in the integrand are known. This general form with arbitrary coupling constants and free integration constants, contains all the potentials we can take in this system and it has not been appeared in the literature before. This general form of potential enables us to obtain the exact solution for metric function (\ref{Metric2}) which gives the relation for metric and potential functions. From (\ref{Metric2}) we can see that once we know the expression of potential the metric function is known.

Since the integral in $U_2(r)$ cannot be integrated analytically, special attention should be given to the continuity of the potential and the metric function for further analysis.

We start our consideration with  negative values of $\delta$. 
For $-1<\delta<0$, $\Bbb{P}(r)$ and $U_1(r)^{-2}$ are both continuous. The potential $U(r)$ is also continuous because $U_1(r)$ is continuous.
For other cases, namely $\delta>0$ and $\delta<-1$, the continuity conditions for $\Bbb{P}(r)$ and $U_1(r)^{-2}$ are identical:
\begin{equation}
\delta A^\delta \leq 4B^\delta (1+\delta) \quad \text{i.e.} \quad 4(1+\delta)\gamma^\delta \geq \delta. \label{continuity}
\end{equation}
From equation (\ref{Metric}), we can see that if the potential is continuous then so is the metric function. Therefore the continuity condition (\ref{continuity}) can ensure the continuity of the potential and the metric.


Then we begin to analyze the asymptotic behavior of the potential at the leading order
\begin{eqnarray}
U(r \to \infty)&=&C_1 U_1(r \to \infty)+C_2 U_2(r \to \infty)\notag\\
&=&C_1 U_1(r \to \infty)+C_2 U_1(r \to\infty)\lim_{r \to \infty}\int U_1(r)^{-2}\Bbb{P}(r)\text{d}r\notag \\
&=&U_1(r \to \infty)\left[C_1+C_2 \lim_{r \to \infty}\int U_1(r)^{-2}\Bbb{P}(r)\text{d}r\right],
\end{eqnarray}
where
\begin{eqnarray}
\lim_{r \to \infty}\int U_1(r)^{-2}\Bbb{P}(r)\text{d}r=\left\{
     \begin{array}{lr}
     \frac{\delta  (\delta +1) r^{-\delta -2}}{4 (\delta +2)} \to 0 & \delta>0 \\
     \frac{32 (\delta +1)^3 A^{-3 \delta } r^{2 \delta -2}}{(\delta -2)^2 (\delta -1) \delta ^2}  \to 0  & -1<\delta<0 \\
     -\frac{32 B (\delta +1)^2 A^{-2 \delta } r^{\delta -3}}{(\delta -3) (\delta -2)^2 (\delta -1) \delta ^2} \to 0 & \delta<-1
\end{array}
\right. .
\end{eqnarray}
So we have
\begin{eqnarray}
U(r \to \infty)=C_1 U_1(r \to \infty)=\left\{
             \begin{array}{lr}
              C_1=\Lambda    &   \delta>0  \\
             -\frac{C_1\delta  (\delta-1)(\delta-2) A^{\delta } }{8 (\delta +1)} r^{-\delta } \to Sgn(C_1)\infty & -1< \delta <0  \\
             -\frac{C_1\delta  (\delta-1)(\delta-2) A^{\delta } }{8 (\delta +1)} r^{-\delta } \to -Sgn(C_1)\infty &  \delta<-1
             \end{array}.
\right. ,
\end{eqnarray}
from which we can see that for positive $\delta$ the constant of integration $C_1$ is the effective cosmological constant $\Lambda=U(\Phi=0)=U(r \to \infty)=C_1$, but for negative $\delta$ the cosmological constant is difficult to to be defined from the knowledge of the potential.
The asymptotic behavior of the potential at $r \to 0$ is,
\begin{eqnarray}
U(r\to 0)&=&C_1 U_1(r\to 0)+C_2 U_2(r\to 0)\notag\\
&=&C_1 U_1(r \to 0)+C_2 U_1(r \to 0)\lim_{r \to 0}\int U_1(r)^{-2}\Bbb{P}(r)\text{d}r\notag \\
&=&U_1(r \to 0)\left[C_1+C_2 \lim_{r \to 0}\int U_1(r)^{-2}\Bbb{P}(r)\text{d}r\right],
\end{eqnarray}
where
\begin{eqnarray}
U_1(r\to 0)=1-\frac{\delta  }{4 (\delta +1)\gamma^\delta}
\end{eqnarray}
is a constant while
\begin{eqnarray}
  \lim_{r \to 0}\int U_1(r)^{-2}\Bbb{P}(r)\text{d}r= \frac{8 (\delta +1)^2 B^{\delta -2} \left((\delta -1) \delta  A^{\delta }+4 (\delta +1)^2 B^{\delta }\right)}{r \left(4 (\delta +1) B^{\delta }-\delta  A^{\delta }\right)^3} \to +\infty
\end{eqnarray}
is going to infinity. Using  the continuity condition $4(1+\delta)\gamma^\delta > \delta$ we finally get
\begin{eqnarray}
U(r\to 0)&=&C_2 U_1(r \to 0) \lim_{r \to 0}\int U_1(r)^{-2}\Bbb{P}(r)\text{d}r \to  \left\{
             \begin{array}{lr}
              Sgn(C_2)\infty  \quad  \quad  & \delta>0  \\
              Sgn(C_2)\infty  \quad \quad  &  -1< \delta <0  \\
              -Sgn(C_2) \infty \quad  \quad  &  \delta<-1
             \end{array}.
\right. ,
\end{eqnarray}
so the potential goes to infinity at both $r \to \infty$ and $r \to 0$. Therefore we will rely on the metric function to find out the asympotic behavior of  the spacetime.

 The asymptotic  behavior of the metric function at $r \to \infty$ and $r\to 0$ is
\begin{eqnarray}
f(r\to \infty)&=&-C_1 r^2=-\Lambda r^2 ,\label{fi}\\
f(r \to 0)&=&-\frac{8 C_2 (\delta +1)^2 \gamma^{\delta }}{\delta ^2 A^{2\delta }\left(\delta  -4 (\delta +1) \gamma^{\delta }\right)}=const., \label{f0}
\end{eqnarray}
from which we can see that for all values of $\delta$ the constant $C_1$ plays the role of cosmological constant.
Under the continuity condition $4(1+\delta)\gamma^\delta > \delta$ , the sign of $f(r \to 0)$ is  $\text{sgn} \left(f(r \to 0)\right)=\text{sgn}(C_2)$.
For $\delta=1$, we recover the BTZ black hole \cite{Banados:1992wn}.

 We now consider the case where the continuity condition  is  $4(1+\delta)\gamma^\delta = \delta$. 
At $r\to 0$ the asymptotic expressions of potential and metric functions are same
\begin{eqnarray}
U(r\to 0)&=&-\frac{C_2 \left(\delta ^2-6 \delta -7\right) B^{-\delta -2}}{72 \delta }=const., \\
f(r\to 0)&=&\frac{C_2 B^{1-\delta }}{3 \delta ^2 r} \to Sgn(C_2)\infty .
\end{eqnarray}
At $r\to \infty$ we have
\begin{eqnarray}
U(r\to \infty)&=&\left\{
             \begin{array}{lr}
              C_1  \quad  \quad  & \delta>0  \\
              -\frac{C_1\delta  (\delta-1)(\delta-2) A^{\delta } }{8 (\delta +1)} r^{-\delta }  \quad \quad  &  -1< \delta <0  \\
              -\frac{C_1\delta  (\delta-1)(\delta-2) A^{\delta } }{8 (\delta +1)} r^{-\delta }  \quad  \quad  &  \delta<-1
             \end{array},
\right. \\
f(r\to \infty)&=&-C_1 r^2=-\Lambda r^2 ,
\end{eqnarray}
Therefore the asymptotic expressions are exactly the same with the case for $4(1+\delta)\gamma^\delta > \delta$.

  Another way to see the properties of the spacetime at large distances is to consider the perturbations of a neutral massless scalar field in the spacetime of a black hole. Consider the equation of the scalar field
$\phi_0$
\begin{eqnarray}
\square \phi_0 &=&0~.
\end{eqnarray}
If we  transform the scalar field as  $\phi_0=r^{-1/2}\varphi_0 e^{-i\omega_0 t}$, then with the use of the tortoise coordinate
$r_*=\int \frac{\text{d}r}{f(r)}$ the scalar equation  can be written in a Schrodinger-like form as
\begin{eqnarray}
\frac{\text{d}^2 \varphi_0}{\text{d}r_*^2}+\left(\omega_0^2-V_\text{eff}\right)\varphi_0=0~,
\end{eqnarray}
where the effective potential in the background of the spherically symmetric metric (\ref{ansatz}) is
\begin{eqnarray}
V_\text{eff}&=&-\frac{f(r)\left(f(r)-2rf'(r)\right)}{4r^2}~.
\end{eqnarray}
The asymptotic behavior of the effective potential is $V_\text{eff}(r\to\infty)=3\Lambda^2 r^2/4$. Therefore,  for all cases and independently of the sign of $\delta$  we have a potential barrier at AdS boundary that can constraint the matter fields.

%
%
%
%
%

Since the integral in $U_2(r)$ cannot be solved while all parameters being free, we need to fix some parameters and confirm our solution by consistency check. We set $\delta=1$, $A=1/q$, $B=1/(8q)$, $C_1=-\frac{1}{\ell^2}$ and $C_2=6144\alpha q$ where the additional free parameter $q$ characterizes the strength of the scalar field.
We find that the potential and metric function become
\begin{eqnarray}
U(r)&=&-\frac{813qr+24q^2r^2+64 q^3 r^3+32\alpha\ell^2}{\ell^2\left(1+8qr\right)^3}~, \\
U(\Phi)&=&-\frac{1}{\ell^2}+\left(\frac{1}{512\ell^2}-\frac{\alpha}{2}\right)\Phi^6~,\\
f(r)&=& \frac{r^2}{\ell^2}-\frac{12\alpha}{q^2}-\frac{\alpha}{q^3 r}~,     \label{Xu}
\end{eqnarray}
which can reduce to the static limit of the black hole solution in \cite{Xu:2014xqa}. This potential form has been widely studied in static exact black hole solutions dressed with non-minimally coupled scalar field \cite{Xu:2014xqa,Martinez:1996gn,Nadalini:2007qi}. In  \cite{Fan:2015tua} a random potential $V$ was introduced but the purpose of that work was  not to find the most general solution for a given scalar potential $V(\varphi)$.

\begin{figure}[h]
\centering
\includegraphics[width=0.55\textwidth]{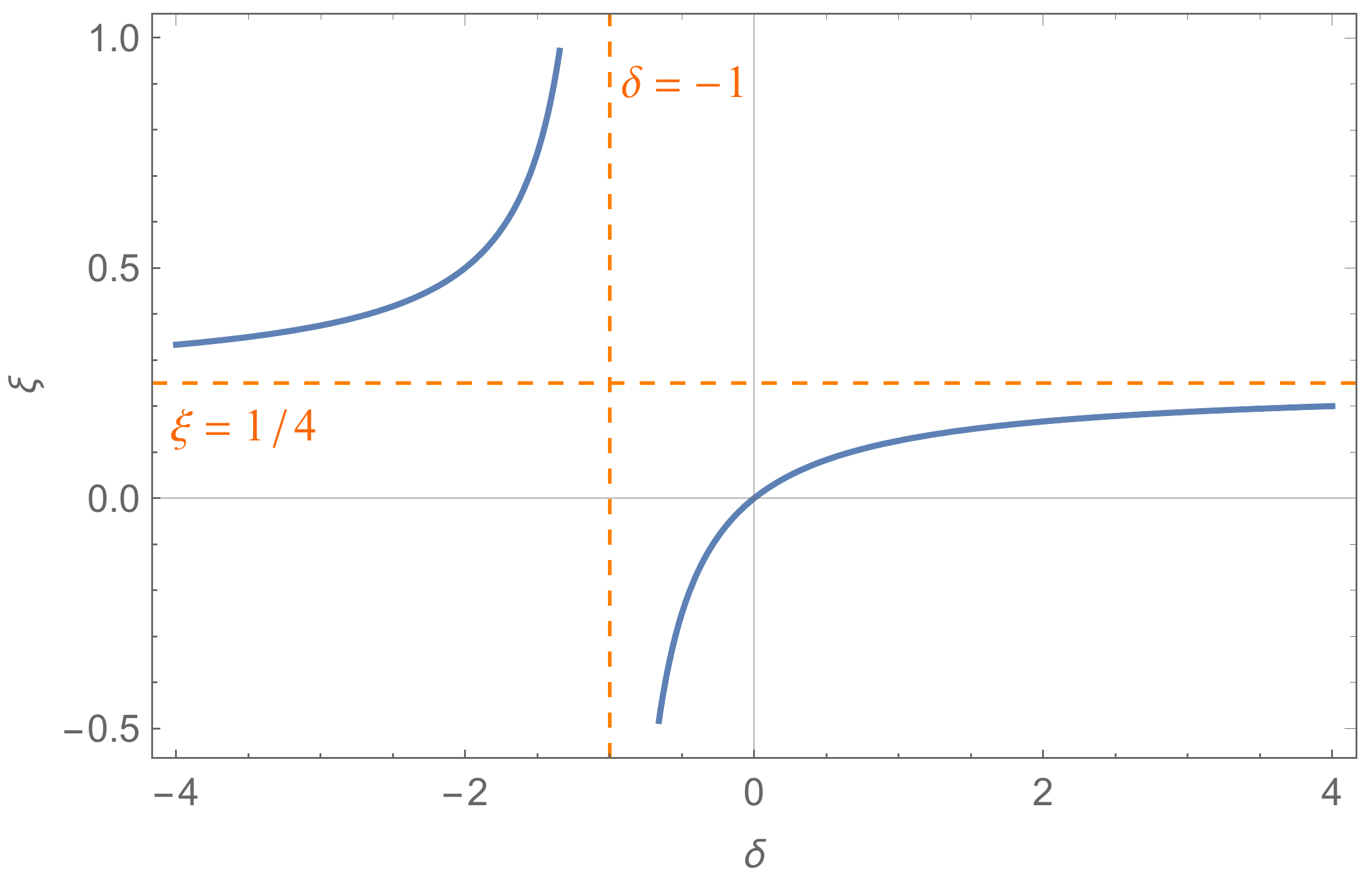}
\caption{The coupling constant $\xi$ always rises with $\delta$. Note that $\delta \neq -1$ and $\xi \neq 1/4$, the former corresponds to $\xi \neq \pm \infty$.  }
\label{fig:xi}
\end{figure}

Let us now discuss the general solution. If $C_1<0$ and $C_2<0$, the relation (\ref{fi}) and (\ref{f0}) can lead to $f(r\to 0)<0$ and $f(r\to \infty)>0$. Then there must be a zero in the metric function, which corresponds to the event horizon of an AdS black hole.
Note that what really has physical meaning is the coupling constant $\xi$ rather than $\delta$, so we first plot the relation $\xi(\delta)=\frac{\delta}{4(1+\delta)}$ in Fig. \ref{fig:xi}.

\begin{figure}[h]
\centering%
\subfigure[~Scalar potential $U(r)$]{
 \includegraphics[width=.45\textwidth]{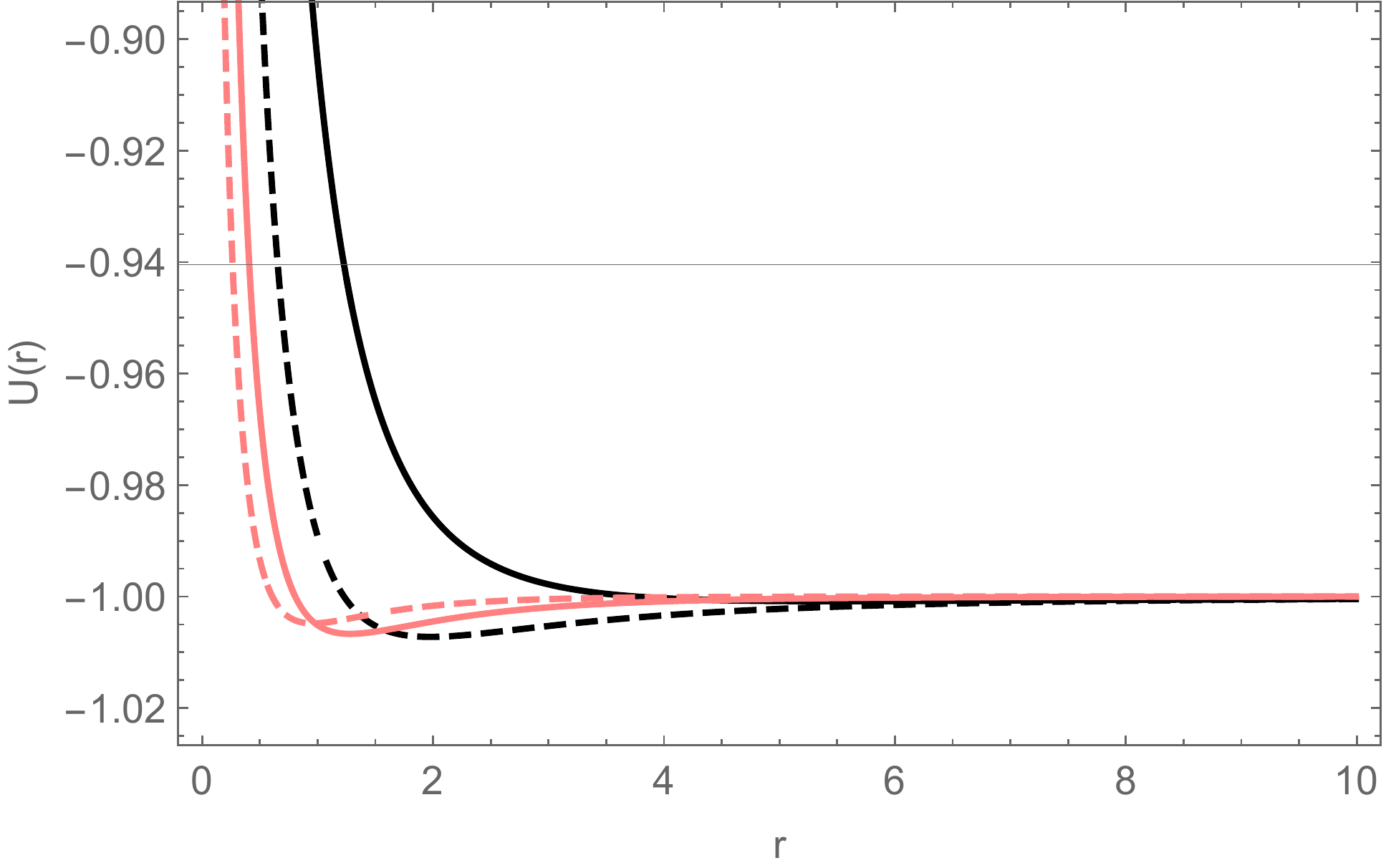} }
\subfigure[~Metric function $f(r)$]{
 \includegraphics[width=.45\textwidth]{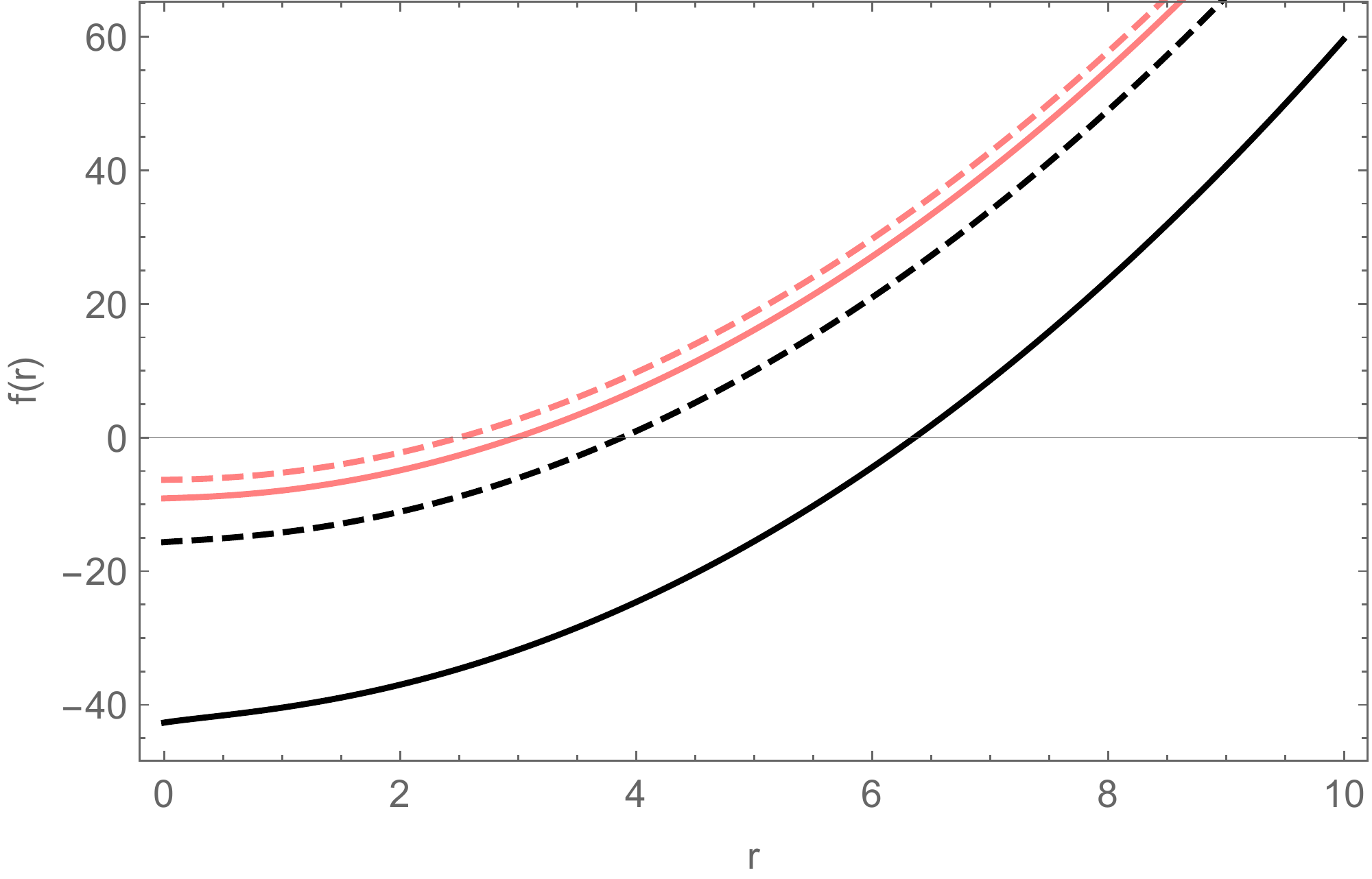} }
\caption{{}The black curve, black dashed curve, pink curve and pink dashed curve correspond to $\delta=1,2,3,4$, i.e. $\xi=1/8,1/6,3/16,1/5$, respectively. Large coupling constant $\xi$ corresponds to black holes with small event horizon when $1/8 \leq \xi<1/4$.}
\label{fig:AdS1}
\end{figure}

\begin{figure}[h]
\centering%
\subfigure[~Scalar potential $U(r)$]{
 \includegraphics[width=.45\textwidth]{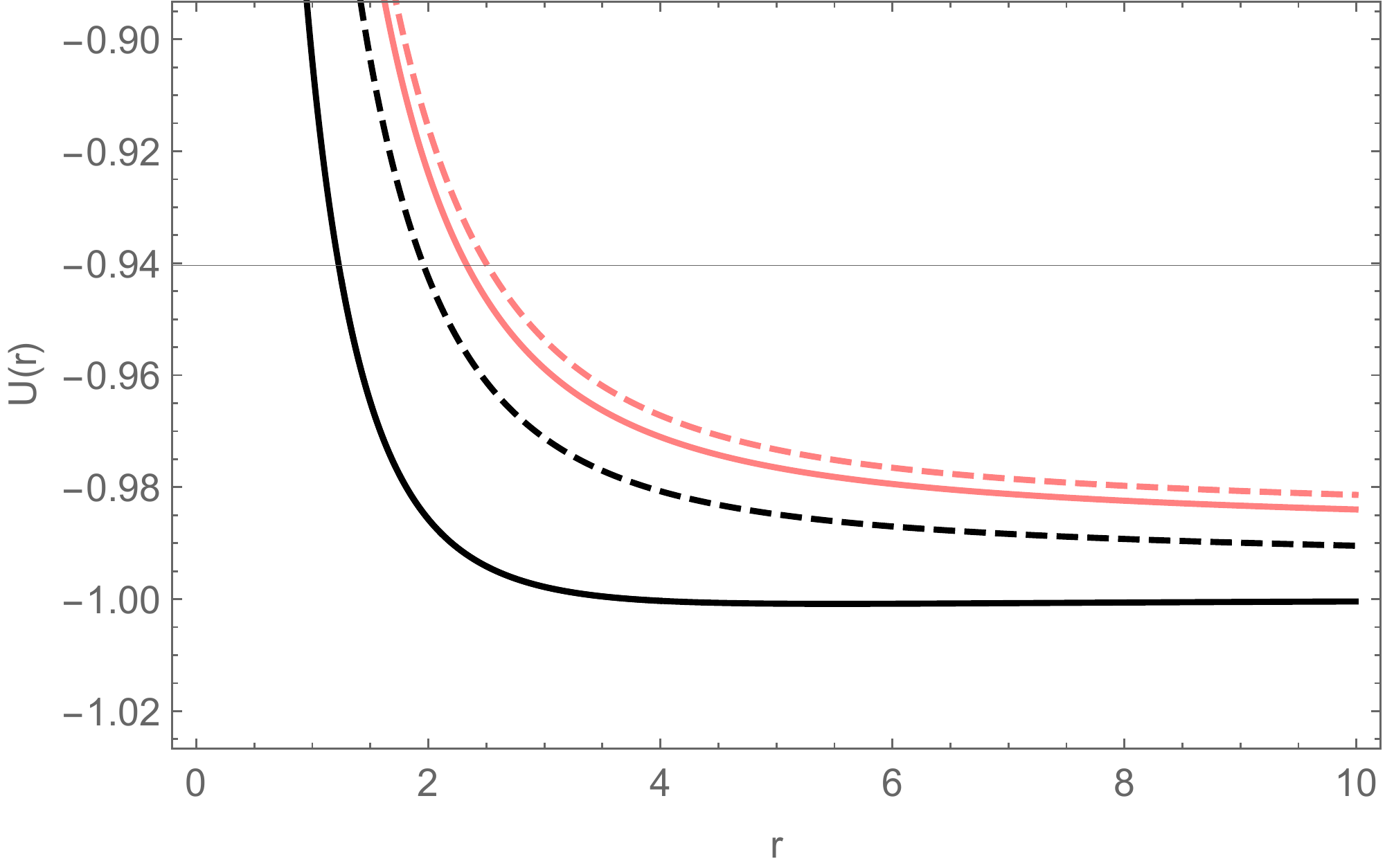} }
\subfigure[~Metric function $f(r)$]{
 \includegraphics[width=.45\textwidth]{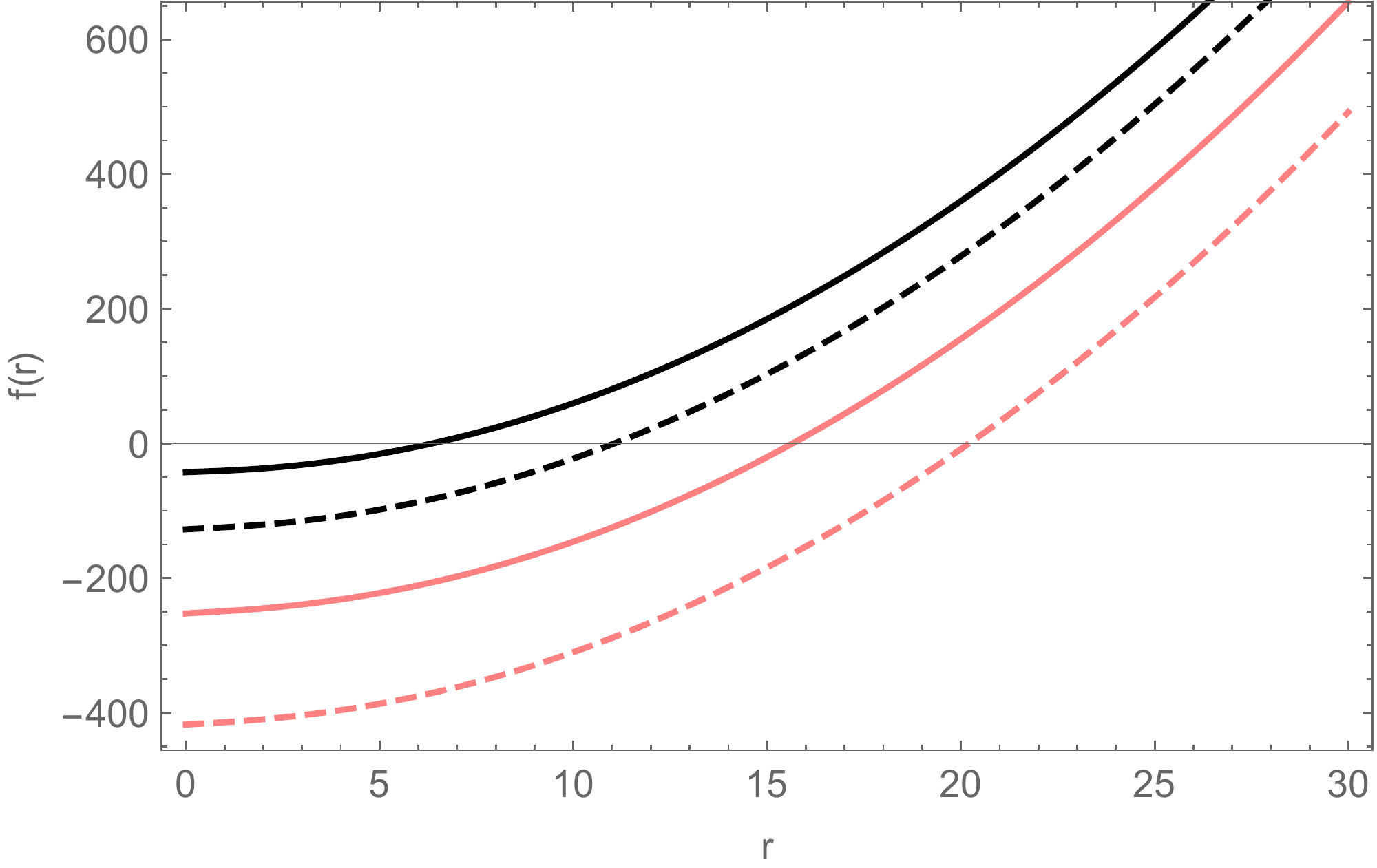} }
\caption{{}The black curve, black dashed curve, pink curve and pink dashed curve correspond to $\delta=1,1/2,1/3,1/4$, i.e. $\xi=1/8, 1/12, 1/16, 1/20$, respectively. Large coupling constant $\xi$ corresponds to black holes with small event horizon when $0<\xi \leq 1/8$. }
\label{fig:AdS2}
\end{figure}

\begin{figure}[h]
\centering%
\subfigure[~Scalar potential $U(r)$]{
 \includegraphics[width=.45\textwidth]{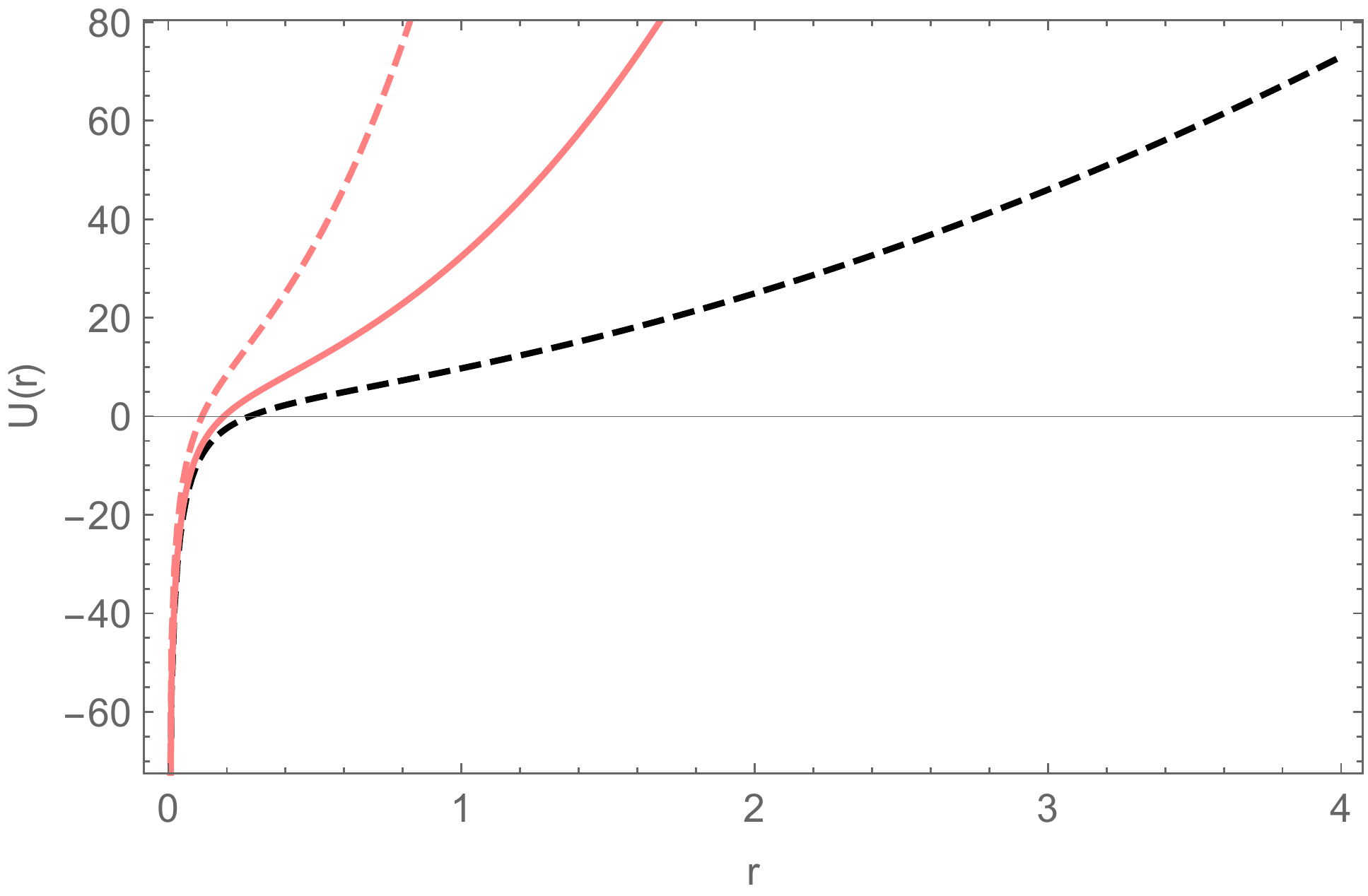} }
\subfigure[~Metric function $f(r)$]{
 \includegraphics[width=.45\textwidth]{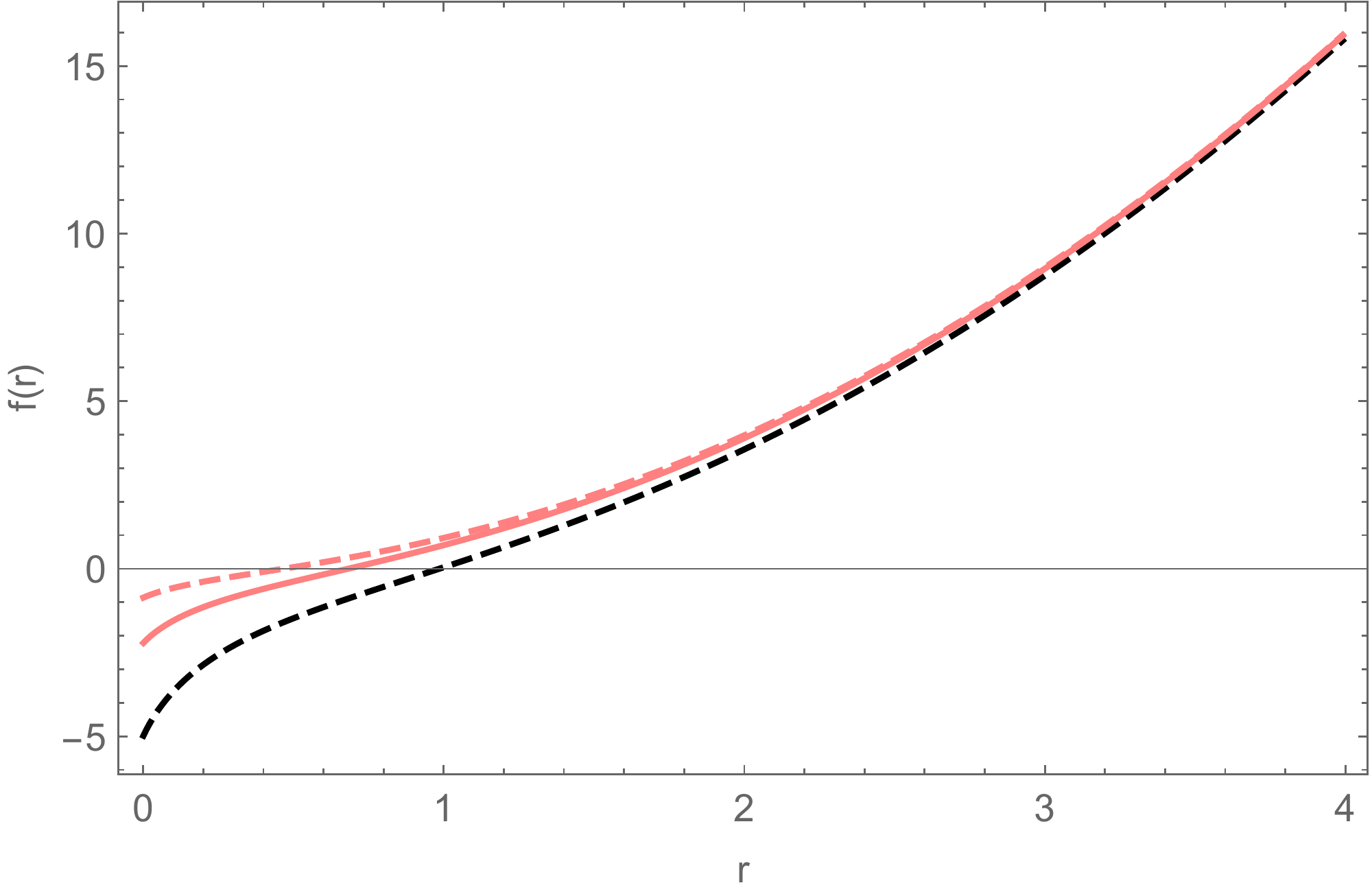} }
\caption{{}The black dashed curve, pink curve and pink dashed curve correspond to $\delta=-2,-3,-4$, i.e., $\xi=1/2,3/8,1/3$ respectively. Large coupling constant $\xi$ corresponds to black holes with large event horizon when $\xi>1/4$.}
\label{fig:AdS3}
\end{figure}

\begin{figure}[h]
\centering%
\subfigure[~Scalar potential $U(r)$]{
 \includegraphics[width=.45\textwidth]{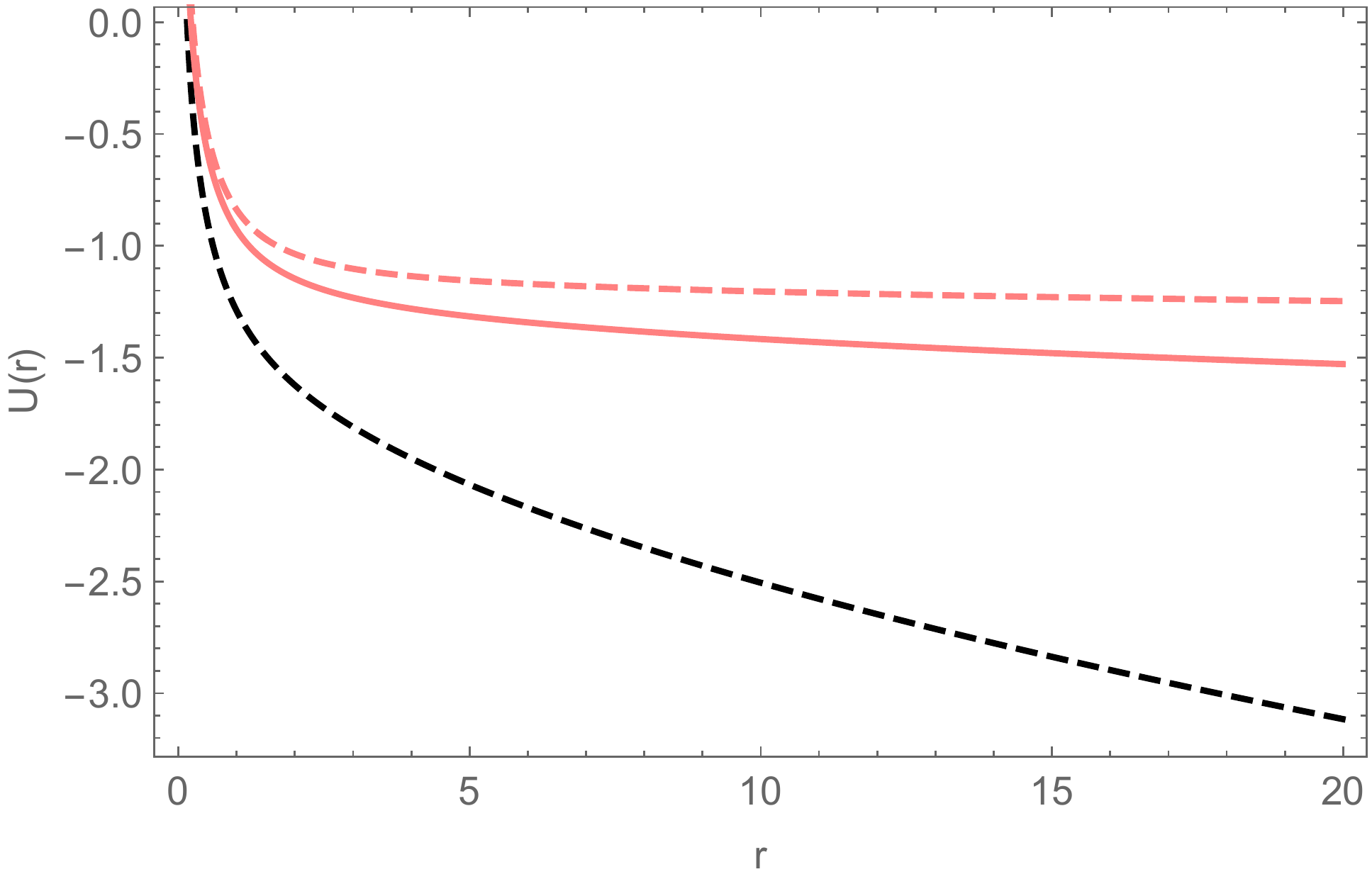} }
\subfigure[~Metric function $f(r)$]{
 \includegraphics[width=.45\textwidth]{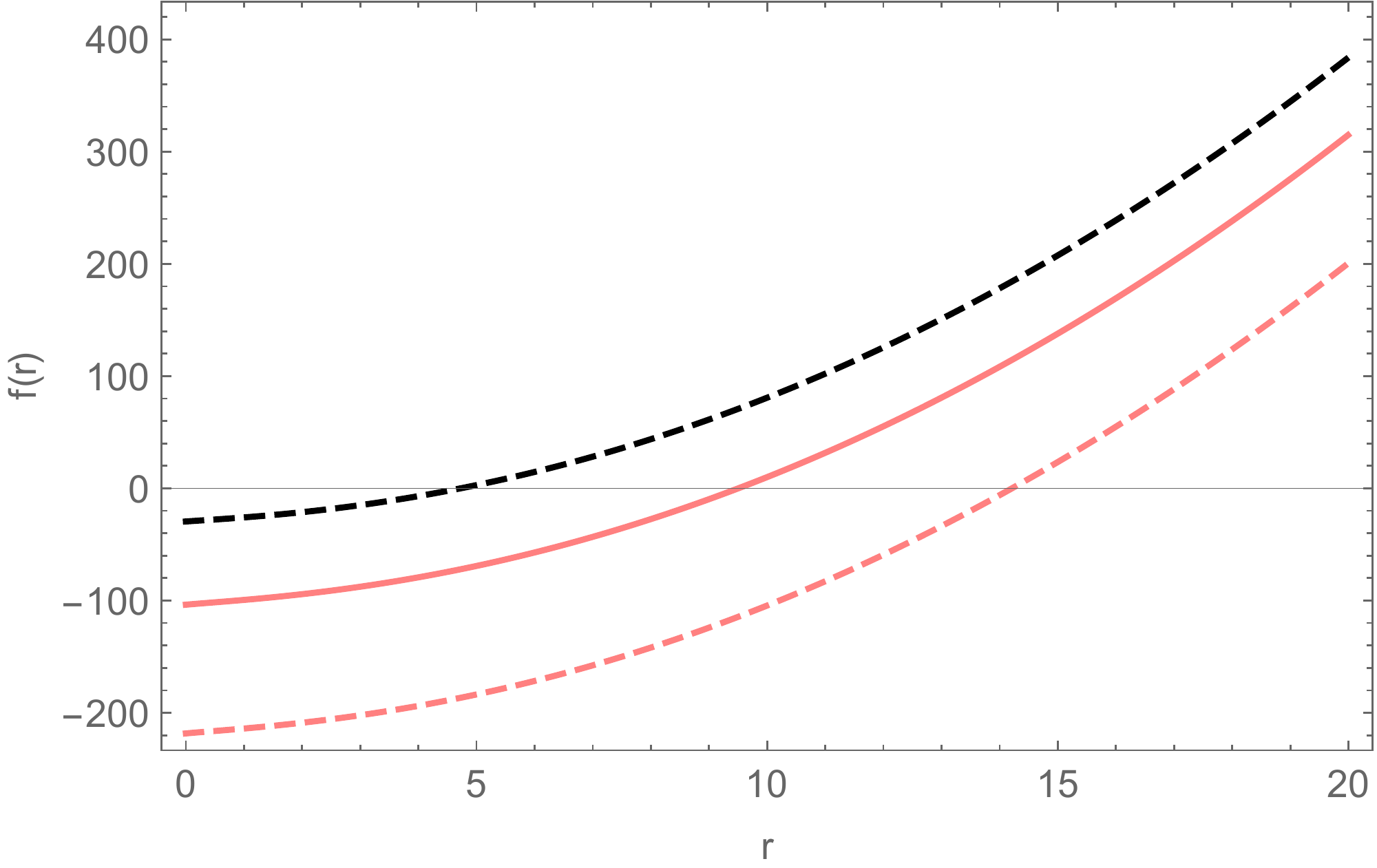} }
\caption{{}The black dashed curve, pink curve and pink dashed curve correspond to $\delta=-1/2,-1/3,-1/4$, i.e. $\xi=-1/4,-1/8,-1/12$ respectively. Large coupling constant $\xi$ corresponds to black holes with large event horizon when $\xi<0$. }
\label{fig:AdS4}
\end{figure}

From the figure we can see that positive $\delta$ corresponds to $0<\xi<1/8$, $\delta<-1$ corresponds to $\xi>1/8$ and $-1<\delta<0$ corresponds to negative $\xi$. We plot the potential and metric functions with different ranges of $\delta$ respectively in Fig. \ref{fig:AdS1}, Fig. \ref{fig:AdS2}, Fig. \ref{fig:AdS3} and Fig. \ref{fig:AdS4}, in all of which we have fixed the other parameters as: $A=1$, $B=2$, $\Lambda=C_1=-1$ and $C_2=-10$. Table \ref{Table:coupling} shows the influence of the coupling constant $\xi$ and $\delta$.

\newpage

\begin{table}
\caption{The influence of the coupling constant $\xi$ and $\delta$}
\centering
\begin{tabular}{|c|c|c|c|}
\hline
$0<\delta \leq 1$ & $\delta \geq 1$ & $-1 <\delta < 1$ & $\delta < -1$ \\
\hline
$0<\xi \leq 1/8$ &  $1/8 \leq \xi < 1/4$ & $ \xi <0$ & $\xi >1/4$ \\
\hline
\multicolumn{2}{|c|}{Large $\xi$ leads to small event horizon.}
& \multicolumn{2}{c|}{Large $\xi$ leads to large event horizon.} \\
\hline
\multicolumn{4}{|c|}{Large absolute values of $\delta$ correspond to small radius of event horizon.}\\
\hline
\end{tabular}\label{Table:coupling}
\end{table}

From the figures we can see that for $0<\xi<1/4$, large coupling constants $\xi$ correspond to black holes with small radius of event horizons while for $\xi<0$ and $\xi>1/4$, large coupling constants $\xi$ correspond to large radius of event horizons. (For $\xi < 0$, to be larger means its absolute value is smaller.)
Also note that this change of behavior corresponds to the sign reversal of $\delta$, so there is  a consistent influence from $\delta$: large absolute values of $\delta$ correspond to black holes with small radius of event horizon, and {the gradient of self-potential of the scalar field becomes steeper at the origin}. Finally, when $\delta \to \pm \infty$, the curve will correspond to $\xi=1/4$. {Note that in all of our plots, we have fixed $B=2$ and $A=1$, i.e., $\gamma=2$. With this value of $\gamma$, arbitrary value of $\delta$ can satisfy the continuity condition $4(1+\delta)\gamma^\delta \geq \delta$.}


Lastly we check for the curvature singularity of the solutions.
The Ricci scalar and the Kretschmann scalar are given by
\begin{equation}
R(r)=-f''(r)-\frac{2f'(r)}{r}~, \qquad \text{and}~~ K(r)=f''(r)^2+\frac{2f'(r)^2}{r^2},
\end{equation}
respectively.
We note that
\begin{eqnarray}
f'(r \to 0)&=&\left\{
     \begin{array}{lr}
     -\frac{24 \left(C_2 (\delta +1)^2 B^{\delta -1} \ln{r}\right)}{\left(\delta  A^{\delta }-4 (\delta +1) B^{\delta }\right)^2}, \qquad ~~  4(1+\delta)\gamma^\delta > \delta    \\
     -\frac{C_2 B^{1-\delta }}{3 \delta ^2 r^2},\qquad \qquad \qquad \qquad   4(1+\delta)\gamma^\delta= \delta
     \end{array}
     \right.\\
f''(r \to 0)&=&\left\{
     \begin{array}{lr}
     -\frac{24 C_2 (\delta +1)^2 B^{\delta -1}}{r \left(\delta  A^{\delta }-4 (\delta +1) B^{\delta }\right)^2}, \qquad \qquad  4(1+\delta)\gamma^\delta > \delta    \\
     \frac{2 C_2 B^{1-\delta }}{3 \delta ^2 r^3}, \qquad \qquad \qquad \qquad  ~ 4(1+\delta)\gamma^\delta= \delta
     \end{array}
     \right.
\end{eqnarray}
Since $f'\left(r \to 0 \right) \to \infty$ and $f''\left(r \to 0 \right) \to \infty$,
there must be a curvature singularity at $r=0$.

\section{Special Exact Solutions}

Since the integral in $U_2(r)$ cannot be solved with $\delta$ unspecified, let us fix $\delta$ to find some special solutions.
In fact, for $\delta=n,-n,1/n,-1/n$ with $n=1,2,3...$, we can always -- in principle -- obtain the corresponding exact black hole solutions, however this becomes more complicated with increasing $n$. So we only present the solutions with $\delta=1$, $\delta=2$, $\delta=-2$, $\delta=1/2$ and $\delta=-1/2$ to illustrate the different properties of spacetimes for $\delta\geq 1$, $\delta<-1$, $-1<\delta<0$ and $0<\delta<1$.

\subsection{$\delta=1$}

When $\delta=1$ the scalar field, the metric and the potential functions are
\begin{eqnarray}
\Phi(r)&=&A(r+B)^{-1/2}~,  \\
f(r)&=&\frac{8 A^2 C_2 (B+2 r)-A^2 C_1 r^2 \left(A^4-16 A^2 B+64 B^2\right)-64 B^2 C_2}{A^2 \left(A^2-8 B\right)^2}\notag \\
&&-\frac{128 A^2 C_2 r^2 }{A^2 \left(A^2-8 B\right)^3}\ln{ \frac{r}{8 (B+r)-A^2}}~,\\
U(\Phi)&=&C_1 U_1(\Phi)+C_2 U_2(\Phi)~,  \\
U_1(\Phi)&=&1-\frac{\gamma ^2 \Phi ^6}{8}~,  \\
U_2(\Phi)&=&\frac{32}{A^3 \gamma  (8 \gamma -1)^3 \left(\Phi ^2-8\right)}\left[16 \gamma ^3 \left(\Phi ^2-4\right) \Phi ^4+2 \gamma  \left(\Phi ^2-8\right) \left(\gamma ^2 \Phi ^6-8\right) \ln{\frac{\Phi ^2-8}{1-\gamma  \Phi ^2}}\right.\notag\\
&&\left.-2 \gamma ^2 \left(\Phi ^6-8 \Phi ^4+32 \Phi ^2+256\right)-\gamma  \left(\Phi ^2-16\right) \Phi ^2-\Phi ^2+8\right]~,
\end{eqnarray}
which agree with the black hole solution presented in \cite{Fan:2015tua} except for a clerical error in that paper.

The continuity condition becomes $A \leq 8B$. In fact this solution only describes the case when $A<8B$. If we choose $A=8B$, then the solution reduces to the solution \cite{Xu:2014xqa} as  mentioned before.

\subsection{$\delta=2$}

We can obtain a new exact black hole solution when $\delta=2$, with
\begin{eqnarray}
\Phi(r)&=&\frac{A}{r+B}~,\\
f(r)&=&\frac{1}{A^2 \left(A^2-6 B^2\right)^3}\left\{-\left(A^2-6 B^2\right) \left[A^6 C_1 r^2-12 A^4 B^2 C_1 r^2+9 A^2 B \left(4 B^3 C_1 r^2+B C_2+4 C_2 r\right)\right. \right. \notag \\
&&\left.-54 B^4 C_2 \right] +9 A^2 C_2 r^2 \left(A^2+18 B^2\right) \ln{\frac{6 r^2}{6 (B+r)^2-A^2}} \notag \\
&&\left.+27 \sqrt{6} A B C_2 r^2 \left(A^2+2 B^2\right) \ln{\frac{\sqrt{6} (B+r)+A}{\sqrt{6} (B+r)-A}}\right\}~,\\
U(\Phi)&=&C_1 U_1(\Phi)+C_2 U_2(\Phi)~,  \\
U_1(\Phi)&=&\frac{1}{6} \left(-3 \gamma ^2 \Phi ^4+2 \gamma  \Phi ^3+6\right)~,  \\
U_2(\Phi)&=&\frac{3 \left(3 \gamma ^2 \Phi ^4-2 \gamma  \Phi ^3-6\right)}{2 A^4 \left(6 \gamma ^2-1\right)^3}\left[\frac{12 \left(6 \gamma ^2-1\right) \left(18 \gamma ^3 \Phi +30 \gamma ^2+\gamma  \Phi -1\right)}{\left(54 \gamma ^2-1\right) \left(\Phi ^2-6\right)}+\frac{\left(6 \gamma ^2-1\right) }{\left(54 \gamma ^2-1\right) \left(3 \gamma ^2 \Phi ^4-2 \gamma  \Phi ^3-6\right)}\right.\notag \\
&&\left(1296 \gamma ^5 \Phi ^3+972 \gamma ^4 \left(\Phi ^2+2\right)+36 \gamma ^3 \Phi  \left(7 \Phi ^2+30\right)-24 \gamma ^2 \left(\Phi ^2-24\right)-6 \gamma  \Phi  \left(\Phi ^2+6\right)+\Phi ^2-6\right)\notag \\
&&\left.+\left(18 \gamma ^2+1\right) \ln{\frac{6-\Phi ^2}{(1-\gamma  \Phi )^2}}-6 \sqrt{6} \left(2 \gamma ^3+\gamma \right) \tanh ^{-1}\left(\frac{\Phi }{\sqrt{6}}\right)\right].
\end{eqnarray}
where $A^2<6B^2$ agrees with our continuity condition (\ref{continuity}). If we choose $B=0$, $A=\sqrt{6}a$, $C_1=-\frac{1}{\ell^2}$ and $C_2=-4\alpha a^4$, it will reduce to the black hole solution \cite{Erices:2017izj} when dimension is three.
When $A^2=6B^2$, the solution can be simplified as
\begin{eqnarray}
\Phi (r)&=&\frac{\sqrt{6} B}{B+r}~,   \\
f(r)&=&\frac{C_2 r^2 }{32 B^4}\ln{\frac{r}{2 B+r}}+\frac{C_2 r}{16 B^3}+\frac{3 C_2}{16 B^2}+\frac{C_2}{12 B r}-C_1 r^2~, \\
U(\Phi)&=&C_1 U_1(\Phi)+C_2 U_2(\Phi)~,  \\
U_1(\Phi)&=&\frac{1}{36} \left(-3 \Phi ^4+2 \sqrt{6} \Phi ^3+36\right)~,  \\
U_2(\Phi)&=&-\left(\left(\Phi ^2-6\right) \left(3 \Phi ^3+\sqrt{6} \Phi ^2+6 \Phi +6 \sqrt{6}\right) \ln{\frac{\sqrt{6}-\Phi }{\Phi +\sqrt{6}}}+6 \left(\sqrt{6} \Phi ^4+2 \Phi ^3-2 \sqrt{6} \Phi ^2-4 \Phi +8 \sqrt{6}\right)\right)\notag\\
&&\frac{\left(-3 \Phi ^4+2 \sqrt{6} \Phi ^3+36\right) \left(\sqrt{6} \Phi ^5-36 \Phi ^4+54 \sqrt{6} \Phi ^3-36 \Phi ^2-216 \sqrt{6} \Phi +648\right)}{32 A^4 \left(\sqrt{6}-\Phi \right)^4 \left(\Phi +\sqrt{6}\right) \left(\sqrt{6} \Phi ^2-18 \Phi -18 \sqrt{6}\right) \left(3 \Phi ^3+\sqrt{6} \Phi ^2+6 \Phi +6 \sqrt{6}\right)}.
\end{eqnarray}

\subsection{$\delta=-2$}

For negative $\delta$, we can obtain an exact solution when $\delta=-2$, with
\begin{eqnarray}
\Phi(r)&=&\frac{r+B}{A}~,\\
f(r)&=&\frac{1}{2 \left(2 A^2-B^2\right)^3}\left(-2 \left(2 A^2-B^2\right) \left(2 A^6 C_2+A^4 \left(-B^2 C_2+4 B C_2 r+4 C_1 r^2\right)-4 A^2 B^2 C_1 r^2 +B^4 C_1 r^2\right)\right. \notag \\
&&\left.+ 2 A^4 C_2 r^2 \left(2 A^2+3 B^2\right) \ln{\frac{r^2}{(B+r)^2-2 A^2}}+\sqrt{2} A^3 B C_2 r^2 \left(6 A^2+B^2\right) \ln{\frac{\sqrt{2} (B+r)-2A}{\sqrt{2} (B+r)-2 A}}\right)~,\\
U(\Phi)&=&C_1 U_1(\Phi)+C_2 U_2(\Phi)~,  \\
U_1(\Phi)&=&-\frac{\gamma ^2}{2}+3 \gamma  \Phi -3 \Phi ^2+1~,\\
U_2(\Phi)&=&\frac{\gamma ^2-6 \gamma  \Phi +6 \Phi ^2-2}{2 \left(\gamma ^2-2\right)^3}\left(\left(3 \gamma ^2+2\right) \ln{\frac{2-\Phi ^2}{(\Phi -\gamma )^2}}-\sqrt{2} \gamma  \left(\gamma ^2+6\right) \tanh ^{-1}\left(\frac{\Phi }{\sqrt{2}}\right)\right.\notag\\
&&\left.+\frac{2 \left(\gamma ^2-2\right) \left(\gamma ^3 \Phi +14 \gamma ^2+22 \gamma  \Phi +20\right)}{\left(\gamma ^2-50\right) \left(\Phi ^2-2\right)}+\frac{18 \left(\gamma ^2-2\right) \left(\gamma ^4-2 \gamma ^3 \Phi -12 \gamma ^2+52 \gamma  \Phi +20\right)}{\left(\gamma ^2-50\right) \left(\gamma ^2-6 \gamma  \Phi +6 \Phi ^2-2\right)}\right)~,
\end{eqnarray}
which is the solution for $2A^2<B^2$.

When $A^2=B^2/2$ we have
\begin{eqnarray}
\Phi (r)&=&\frac{\sqrt{2} (B+r)}{B}~,\\
f(r)&=&\frac{8 B^3 C_2-6 B^2 C_2 r+3 C_2 r^3 \ln{ \frac{r}{2 B+r}}+6 B C_2 r^2-96 C_1 r^3}{96 r}~,\\
U(\Phi)&=&C_1 U_1(\Phi)+C_2 U_2(\Phi)~,  \\
U_1(\Phi)&=&3 \left(\sqrt{2}-\Phi \right) \Phi~,\\
U_2(\Phi)&=&\frac{\left(\sqrt{2} \Phi ^3-6 \Phi ^2+6 \sqrt{2} \Phi -4\right) \left(4 \left(3 \Phi ^2-4\right)+3 \sqrt{2} \Phi  \left(\Phi ^2-2\right) \ln{\frac{\sqrt{2}-\Phi }{\Phi +\sqrt{2}}}\right)}{64 \left(\sqrt{2}-\Phi \right)^2 \left(\Phi ^2-2\right)}~.
\end{eqnarray}

\subsection{$\delta=1/2$}

We can also obtain an exact black hole solution with $\delta=1/2$, with
\begin{eqnarray}
\Phi(r)&=&\left(\frac{A}{r+B}\right)^{1/4}~,\\
f(r)&=&\frac{1}{\sqrt{A} B^{3/2} (A-2^4 3^2 B)^3}\left\{\sqrt{B} (2^4 3^2 B-A) \left[72 A^{3/2} \left(-4 B^2 C_1 r^2+C_2 r \sqrt{B+r}+2 B C_2 \sqrt{B+r}\right) \right. \right. \notag \\
&&+A^{5/2} B C_1 r^2+2^7 3^4 \sqrt{A} B \left(2 B^2 C_1 r^2+3 C_2 r \sqrt{B+r}-2 B C_2 \sqrt{B+r}\right)+2^6 3^3 A B C_2 (B+2 r) \notag \\
&&\left.-2^{10} 3^5 B^3 C_2\right]+36 \left[\sqrt{A} C_2 r^2 \left(A^2-2^5 3^3 A B-2^8 3^5 B^2\right) \ln{\frac{2 \sqrt{B} \sqrt{B+r}+2 B+r}{r}} \right. \notag \\
&&\left.\left.+2^9 3^3 A B^{3/2}C_2 r^2 \ln{\frac{r}{2^4 3^2 (B+r)-A}}+2^9 3^3 A B^{3/2} C_2 r^2 \ln{\frac{2^4 3^2 \left(\sqrt{A}+12 \sqrt{B+r}\right)}{12 \sqrt{B+r}-\sqrt{A}}}\right]\right\}~, \\
U(\Phi)&=&C_1 U_1(\Phi)+C_2 U_2(\Phi)~,  \\
U_1(\Phi)&=&\frac{1}{96} \left(-3 \gamma ^2 \Phi ^{10}-2 \gamma  \Phi ^6-3 \Phi ^2+96\right)~,\\
U_2(\Phi)&=&-\frac{3 \left(3 \gamma ^2 \Phi ^{10}+2 \gamma  \Phi ^6+3 \Phi ^2-96\right)}{8 A^{5/2} (2^4 3^2 \gamma -1)^3}\left[\frac{\left(2^8 3^5 \gamma ^2+2^5 3^3 \gamma -1\right) \tanh ^{-1}\left(\sqrt{\gamma } \Phi ^2\right)}{\gamma ^{3/2}}-2^8 3^3 \ln{\frac{1-\gamma  \Phi ^4}{\left(\Phi ^2-12\right)^2}}\notag \right.\\
&&+\frac{2^{11}3^4 (2^4 3^2 \gamma -1)}{(2^4 3^3 \gamma +5) \left(\Phi ^2-12\right)}+\frac{2^4 3^3 \gamma -3}{\gamma  (2^4 3^3 \gamma +5) \left(3 \gamma ^2 \Phi ^{10}+2 \gamma  \Phi ^6+3 \Phi ^2-2^5 3\right)}\left(2^8 3^4 \gamma ^3 \left(7 \Phi ^2+48\right) \Phi ^6 \right.\notag\\
&&+2^5 3^2 \gamma ^2 \left(11 \Phi ^8+120 \Phi ^6+2^3 3^2 23\Phi ^4+2^6 3^5 \Phi ^2+2^{11} 3^4\right)\notag\\
&&\left.\left.+\gamma  \left(-5 \Phi ^8+2^5 3^2 11 \Phi ^4+2^9 3^2 7 \Phi ^2+2^{14} 3^3\right)-5 \left(\Phi ^4-8 \Phi ^2-2^8 3\right)\right)\right]
\end{eqnarray}
where $\sqrt{A}<12\sqrt{B}$ agrees with the continuity condition (\ref{continuity}).

For $\sqrt{A}=12\sqrt{B}$ the solution becomes much simpler
\begin{eqnarray}
\Phi(r)&=&2 \sqrt{3} \left(\frac{B}{B+r}\right)^{1/4}~,  \\
f(r)&=&\frac{C_2 r^2 \ln{\frac{2 \sqrt{B} \sqrt{B+r}+2 B+r}{r}}}{8 B^{5/2}}+\left(\frac{1 }{6 B}-\frac{ r }{4 B^2}+\frac{2 }{3 r}\right)C_2 \sqrt{B+r}+\frac{C_2}{\sqrt{B}}+\frac{2 \sqrt{B} C_2}{3 r}-C_1 r^2~,  \\
U(\Phi)&=&C_1 U_1(\Phi)+C_2 U_2(\Phi)~,  \\
U_1(\Phi)&=&-\frac{\Phi ^{10}}{2^{13} 3^4}-\frac{\Phi ^6}{2^8 3^3}-\frac{\Phi ^2}{2^5}+1~,  \\
U_2(\Phi)&=&-\frac{3\left(\Phi ^{10}+2^5 3^1 \Phi ^6+2^8 3^4 \Phi ^2-2^{13} 3^4\right) \ln{\frac{12-\Phi ^2}{\Phi ^2+12}}+2^3 3^2 \left(\Phi ^8+2^4 3^2 \Phi ^4-2^8 3^4 \Phi ^2-2^{13} 3^2\right)}{64 A^{5/2}}~.
\end{eqnarray}

\subsection{$\delta=-1/2$}

To show the effects for negative $\xi$ we present the solution with $\delta=-1/2$ as well,
\begin{eqnarray}
\Phi(r)&=&\left(\frac{r+B}{A}\right)^{1/4}~, \\
f(r)&=&\frac{1}{B^{3/2} (B-16 A)^3}\left\{    -\sqrt{B}B-16 A\left(64 A^{3/2} B C_2 (B-2 r)-1024 A^{5/2} B C_2-8 A B \right. \right. \notag \\
&&\left(4 B C_1 r^2-3 C_2 r \sqrt{B+r}+2 B C_2 \sqrt{B+r}\right)\left.+128 A^2 \left(2 B \left(C_2 \sqrt{B+r}+C_1 r^2\right)+C_2 r \sqrt{B+r}\right)\right) \notag \\
&&+B^3 C_1 r^2 +128 A^{3/2} B^{3/2} \text{C2} r^2 \ln{\frac{r}{8 \sqrt{A} \sqrt{B+r}+16 A+B+r}} \notag \\
&&\left.+4 A C_2 r^2 \left(-256 A^2+96 A B+3 B^2\right) \ln{\frac{2 \sqrt{B} \sqrt{B+r}+2 B+r}{r}}  \right\}~, \\
U(\Phi)&=&C_1 U_1(\Phi)+C_2 U_2(\Phi)~,  \\
U_1(\Phi)&=&-\frac{\gamma ^2+6 \gamma  \Phi ^4-15 \Phi ^8-32 \Phi ^6}{32 \Phi ^6}~, \\
U_2(\Phi)&=&-\frac{\gamma ^2+6 \gamma  \Phi ^4-15 \Phi ^8-32 \Phi ^6}{4 A^{3/2} (\gamma -16)^3 \Phi ^6}\left\{-\frac{\left(3 \gamma ^2+96 \gamma -256\right) \tanh ^{-1}\left(\frac{\Phi ^2}{\sqrt{\gamma }}\right)}{\gamma ^{3/2}}-16 \ln{\frac{\Phi ^4-\gamma }{\left(\Phi ^2+4\right)^2}}\right.\notag\\
&&+\frac{128 (\gamma -16)}{(\gamma +112) \left(\Phi ^2+4\right)}+\frac{\gamma -16}{\gamma  (\gamma +112) \left(\gamma ^2+6 \gamma  \Phi ^4-15 \Phi ^8-32 \Phi ^6\right)}\notag\\
&&\left(\gamma ^3 \left(3 \Phi ^2-8\right)+\gamma ^2 \left(-45 \Phi ^6+144 \Phi ^4-544 \Phi ^2+2560\right)-32 \gamma  \left(105 \Phi ^6-376 \Phi ^4-56 \Phi ^2+448\right)\right.\notag\\
&&\left.\left.-1792 \left(15 \Phi ^2+32\right) \Phi ^4\right)\right\}~,
\end{eqnarray}
where $16A \neq B$. When $-\delta A^\delta=4 B^\delta(1+\delta)$, i.e. $16A=B$, the solution is simpler:
\begin{eqnarray}
\Phi (r)&=&2 \left(\frac{B}{B+r}\right)^{-1/4}~,\\
f(r)&=&-\frac{C_2 \left(8 B^{3/2}+\left(\frac{3 r^2 }{B}-8 B-2 r \right)\sqrt{B+r}\right)}{12 r}+\frac{C_2 r^2 \ln{\frac{2 \sqrt{B} \sqrt{B+r}+2 B+r}{r}}}{8 B^{3/2}}-C_1 r^2~,\\
U(\Phi)&=&C_1 U_1(\Phi)+C_2 U_2(\Phi)~,  \\
U_1(\Phi)&=&-\frac{8}{\Phi ^6}+\frac{15 \Phi ^2}{32}-\frac{3}{\Phi ^2}+1~,\\
U_2(\Phi)&=&\frac{8 \left(45 \Phi ^6+96 \Phi ^4-48 \Phi ^2+512\right)+3 \left(15 \Phi ^8+32 \Phi ^6-96 \Phi ^4-256\right) \ln{\frac{4-\Phi ^2}{\Phi ^2+4}}}{2^{14}3^1 A^{3/2} \Phi ^6}~.
\end{eqnarray}

\section{de Sitter  black hole solutions in (2+1)-dimensions}


To study the possibility of de Sitter black hole, we study the zeroes of the metric function.
The equivalent equation for $f(r_h)=0$ is $a(r_h)U(r_h)+b(r_h)U'(r_h)=0$, where the roots $r_h$  correspond to the horizons. After substituting in the concrete expressions we obtain a rather complicated equation

\begin{eqnarray}
&&R(r_h)\equiv \int^{r_h} U_1(r)^{-2}\Bbb{P}(r)\text{d}r-\left\{16 C_2 (\delta +1)^2 r_h^2 \left(B+r_h\right)^{\delta } \left(A^\delta (\delta -1) \delta +4 (\delta +1) \left(B+r_h\right)^{\delta }\right)\right. \notag \\
&&+2 B^2\left(4 (\delta +1) \left(B+r_h\right)^{\delta }-A^\delta \delta \right) \left[A^\delta C_1 \delta ^2 r_h^2 \left(A^\delta \delta -4 (\delta +1) \left(B+r_h\right)^{\delta }\right)+8 C_2 (\delta +1)^2 \left(B+r_h\right)^{\delta }\right] \notag \\
&&-A^\delta C_1 \delta ^2 r_h^4 \left(A^4 \delta ^2 \left(\delta ^2-3 \delta +2\right)-4 A^\delta \delta  \left(\delta ^3-2 \delta ^2+\delta +4\right) \left(B+r_h\right)^{\delta }+32 (\delta +1)^2 \left(B+r_h\right)^{2 \delta }\right) \notag \\
&&-4 B r_h \left[A^\delta C_1 \delta ^2 r_h^2 \left(-A^4 (\delta -1) \delta ^2+4 A^\delta \delta  \left(\delta ^2-\delta -2\right) \left(B+r_h\right)^{\delta }+16 (\delta +1)^2 \left(B+r_h\right)^{2 \delta }\right)\right. \notag \\
&&\left. \left. -4 C_2 (\delta +1)^2 \left(B+r_h\right)^{\delta } \left(A^\delta (\delta -2) \delta +8 (\delta +1) \left(B+r_h\right)^{\delta }\right)\right]\right\} \left/ \left\{ A^\delta C_2 \delta ^2 r_h^2 \left(4 (\delta +1) (B+r_h)^{\delta }-A^\delta \delta \right) \right.\right.\notag \\
&& \left[ B^2 \left(8 (\delta +1) (B+r_h)^{\delta }-2 A^\delta \delta \right)+{r_h}^2 \left(8 (\delta +1) (B+r_h)^{\delta }-A^\delta \delta  \left(\delta ^2-3 \delta +2\right)\right)\right. \notag \\
&& \left. \left.+4 B r_h \left(A^\delta (\delta -1) \delta +4 (\delta +1) (B+r_h)^{\delta} \right)\right] \right\}=0~.
\end{eqnarray}
where we have defined the root function $R(r_h)$. It monotonically increases for positive $r_h$:
\begin{equation}
R'(r_h)=\frac{16 (\delta +1)^2 (B+r_h)^{\delta }}{A^\delta \delta ^2 r_h^3 \left(4 (\delta +1) (B+r_h)^{\delta }-A^\delta \delta \right)}>0,~~ R(\infty)=\frac{C_1}{C_2}~, \label{RtFn}
\end{equation}
which is valid for all values of $\delta$. If $C_1$ and $C_2$ have the same signs, then $R(\infty)>0$ means there is one horizon. While if $C_1$ and $C_2$ have opposite signs, then $R(\infty)<0$ means there is no horizon. If we want an dS black hole, at least two horizons are required, so there can not exist dS black holes.

%

If we consider negative $A$ and $B$, the scalar field is restricted to the  region $0\leq r<-B$.
Because of the  continuity condition $4(1+\delta)\gamma^\delta \geq \delta$, the function $R(r_h)$ is divided into two monotonic pieces by the line $r_h=-B$. For $0 \leq r<-B$, there is at most one horizon, but no asymptotically de Sitter black hole.
%

If we consider a metric function containing divergent points then the continuity condition is not valid and in this case the metric function is divided into three parts by the two divergent points (we only consider the non-negative $r_h$)
\begin{equation}
r_h=\pm \left[\frac{\delta}{4(\delta+1)}\right]^{1/\delta}-B, r_h=-B~.
\end{equation}
The function $R(r_h)$ is still monotonic in the region without divergence but there is still no dS black hole.

On the other hand,
an asymptotically flat spacetime requires $C_1=0$, $U(r)=C_2 U_2(r)$ and the metric function becomes
\begin{equation}
f(r)=\frac{C_2(r+B)^{1+\delta}\left[a(r)U_2(r)+b(r)U_2'(r)\right]}{A^\delta\delta\left[\delta r-(r+B)\right]+4(1+\delta)(r+B)^{1+\delta}}~.
\end{equation}
At infinity we have
\begin{eqnarray}
U(r \to \infty)=-\frac{C_2}{(\delta+2)r^{\delta+2}}~,  \\
f_\infty \equiv  f(r \to \infty)=\frac{2 C_2 (\delta +1)}{A^\delta \delta ^2}~,
\end{eqnarray}
which imply $\text{sgn}\left(f(r \to 0)\right)=\text{sgn}(C_2)=\text{sgn}(f_\infty)$.

In this case, the function $R(r_h)$ becomes simpler
\begin{eqnarray}
R(r_h)&=&\int^{r_h} U_1(r)^{-2}IP(r)dr-16 (\delta +1)^2 \left(B+r_h\right)^{\delta +1} \left[-A^\delta B \delta +A^\delta (\delta -1) \delta r_h \right. \notag \\
&&\left. +4 B (\delta +1) \left(B+r_h\right){}^{\delta }+4 (\delta +1) r_h \left(B+r_h\right)^{\delta }\right] \left/
\left\{ A^\delta \delta ^2 r_h^2 \left(4 (\delta +1) \left(B+r_h\right){}^{\delta }-A^\delta \delta \right)\right. \right.\notag \\
&&  \left[B^2 \left(8 (\delta +1) \left(B+r_h\right){}^{\delta }-2 A^\delta \delta \right)+r_h^2 \left(8 (\delta +1) \left(B+r_h\right){}^{\delta }-A^\delta \delta  \left(\delta ^2-3 \delta +2\right)\right) \right. \notag \\
&&\left.\left.  +4 B r_h \left(A^\delta (\delta -1) \delta +4 (\delta +1) \left(B+r_h\right)^{\delta }\right)\right] \right\}~,
\end{eqnarray}
whose derivative is the same as Eq.(\ref{RtFn}).
\begin{eqnarray}
R'(r_h)=\frac{16 (\delta +1)^2 (B+r_h)^{\delta }}{A^\delta \delta ^2 r_h^3 \left(4 (\delta +1) (B+r_h)^{\delta }-A^\delta \delta \right)}>0~,\,\,\,
\, R(\infty)=0~,
\end{eqnarray}
We observe that we have neither a cosmological horizon nor a black hole horizon. When $C_2>0$ the spacetime  is asymptotically flat without black hole solutions.

Therefore, we conclude that there is only asymptotically AdS (2+1)-dimensional black hole in non-minimally coupled theory without electromagnetic field, at least under our metric ansatz assumption.

\subsection{Charged de Sitter black hole solution}

In \cite{Xu:2013nia}, under the action
\begin{eqnarray}
I=\frac{1}{2}\int d^3 x \sqrt{-g}\left[R-g^{\mu\nu} \nabla_\mu\phi \nabla_\nu\phi-\xi R \Phi^2-2V\left(\phi\right) -\frac{1}{4}F_{\mu\nu}F^{\mu\nu}\right],
\end{eqnarray}
a charged scalar black hole solution is obtained
\begin{eqnarray}
\phi(r)&=&\pm \sqrt{\frac{8B}{r+B}}~, \\
f(r)&=&\left(3\beta-\frac{Q^2}{4}\right)+\left(2\beta-\frac{Q^2}{9}\right)\frac{B}{r}-Q^2\left(\frac{1}{2}+\frac{B}{3r}\right)\ln{r}+\frac{r^2}{\ell^2}~, \\
V(\phi)&=&-\frac{1}{\ell^2}+\frac{1}{512}\left(\frac{1}{\ell^2}+\frac{\beta}{B^2}\right)\phi^6-\frac{Q^2}{18432B^2}\left(192\phi^2+48\phi^4+5\phi^6\right) \notag \\
&&+\frac{Q^2}{3B^2}\left[\frac{2\phi^2}{\left(8-\phi^2\right)^2}-\frac{1}{1024}\phi^6\ln{\frac{B\left(8-\phi^2\right)}{\phi^2}}\right]~,
\end{eqnarray}
where $\beta$ is an integral constant and the maxwell field is $A_\mu dx^\mu=-Q \ln{\left(\frac{r}{r_0}\right)}dt$.

We will discuss the possibility of obtaining asymptotically de Sitter black hole solution from this metric. Here $\Lambda=-\ell^{-2}$ appears in $V(\phi)$ as a constant term, which plays the role of the cosmological  constant. In \cite{Xu:2013nia} only negative $\Lambda$ is considered, and so $\Lambda$ is set to be $-\ell^{-2}$. However, the constant $\Lambda$ can either be positive, zero or negative. So we rewrite their solution with general $\Lambda$,
\begin{eqnarray}
\phi(r)&=&\pm \sqrt{\frac{8B}{r+B}}~, \\
f(r)&=&\left(3\beta-\frac{Q^2}{4}\right)+\left(2\beta-\frac{Q^2}{9}\right)\frac{B}{r}-Q^2\left(\frac{1}{2}+\frac{B}{3r}\right)\ln{r}-\Lambda r^2~, \\
V(\phi)&=&\Lambda+\frac{1}{512}\left(-\Lambda+\frac{\beta}{B^2}\right)\phi^6-\frac{Q^2}{18432B^2}\left(192\phi^2+48\phi^4+5\phi^6\right) \notag\\
&&+\frac{Q^2}{3B^2}\left[\frac{2\phi^2}{\left(8-\phi^2\right)^2}-\frac{1}{1024}\phi^6\ln{\frac{B\left(8-\phi^2\right)}{\phi^2}}\right]~.
\end{eqnarray}

\begin{figure}[h]
\centering
\subfigure[Scalar field $\phi(r)$]{
 \label{fig2a}\includegraphics[width=.45\textwidth]{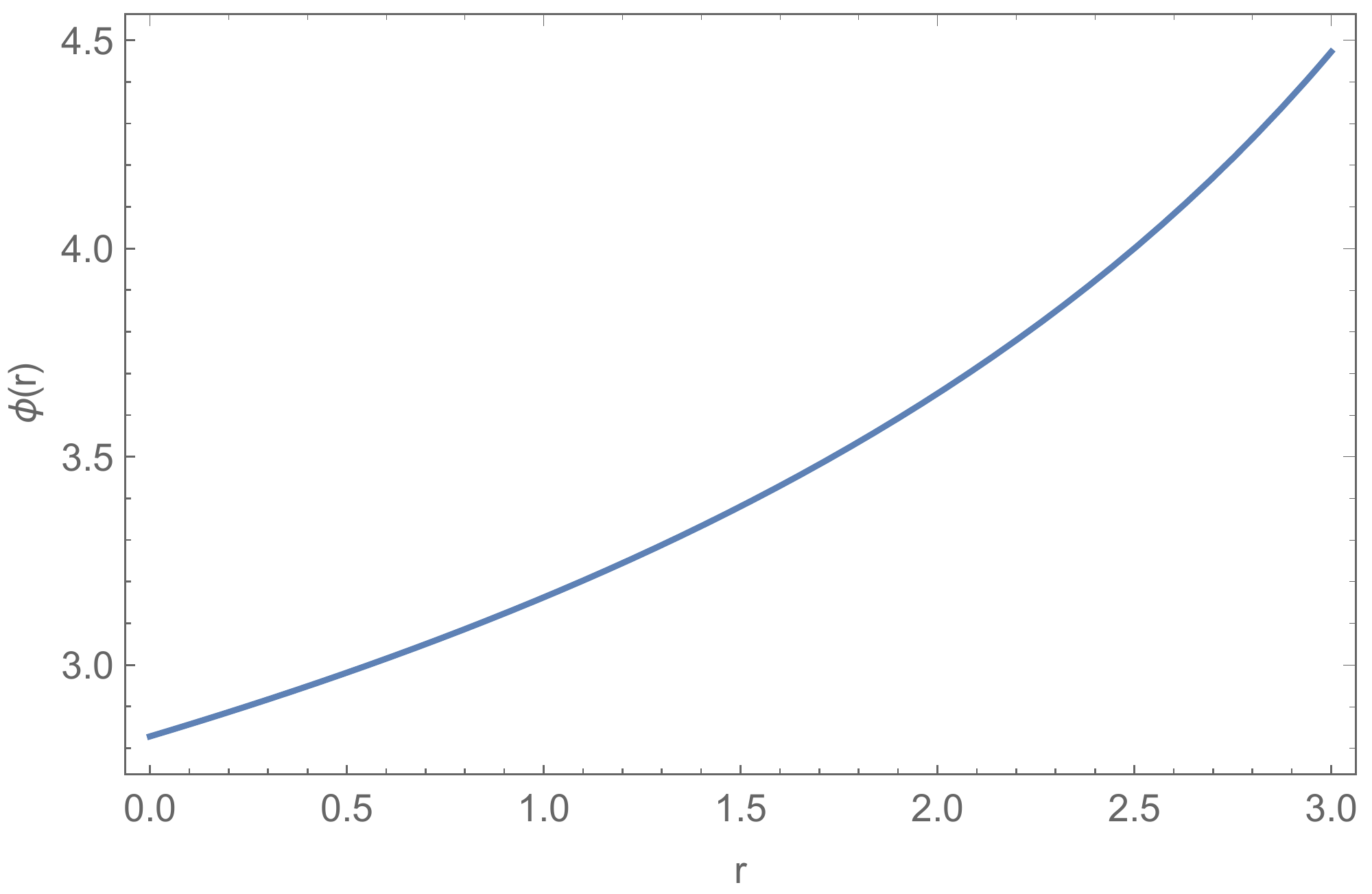} }
\subfigure[Metric function $f(r)$]{
 \label{fig2b}\includegraphics[width=.45\textwidth]{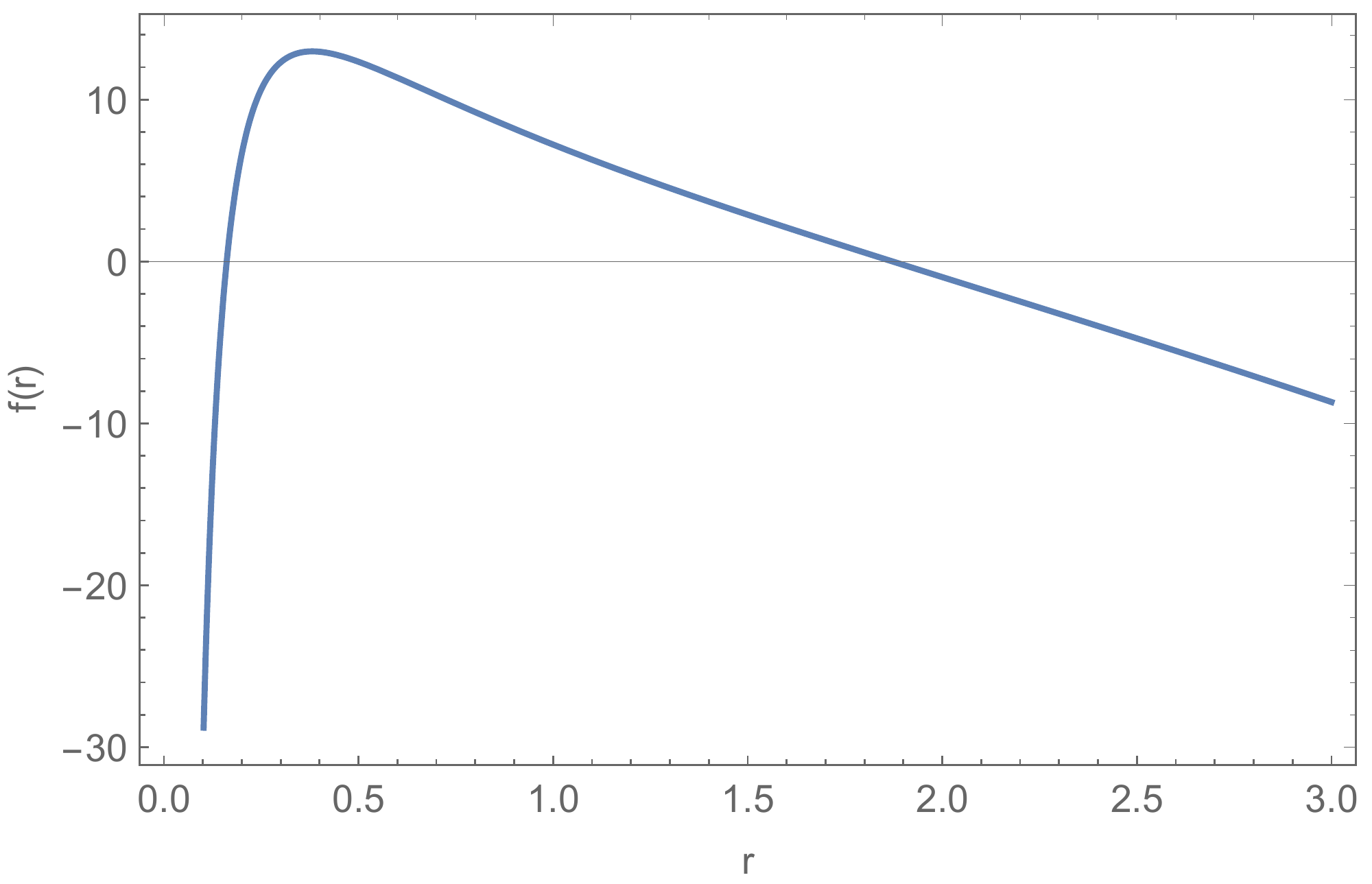} }
\caption{A (2+1)-dimensional asymptotically de Sitter charged black hole. In this example we take $\Lambda=1$, $B=-5$, $Q=2$ and $\beta=-1$.}
\label{fig:dSBH}
\end{figure}

If $B>0$, then there is no black hole for non-negative $\Lambda$. However, once  $B<0$, it is possible to have a black hole solution with positive $\Lambda $, which is a de Sitter black hole solution. The scalar field and the metric functions are shown in Fig. \ref{fig:dSBH}. Note that the scalar field is divergent at $r=-B=5$, but this does not matter because it lies outside the cosmological horizon. The metric function and the curvature are both not divergent at this point.

To check the stability of the spacetime, we study the perturbations of a massless scalar field $\phi_1$ and a massive scalar field $\phi_2$ respectively,
\begin{eqnarray}
\square \phi_1 &=&0,\\
\square \phi_2 &=& m^2 \phi_2,
\end{eqnarray}
where $m$ is the mass of the scalar field $\phi_2$.
The transformations $\phi_1=r^{-1/2}\varphi_1 e^{-i\omega_1 t}$ and $\phi_2=r^{-1/2}\varphi_2 e^{-i\omega_2 t}$ lead to the differential equations
\begin{eqnarray}
\frac{\text{d}^2 \varphi_1}{\text{d}r_*^2}+\left(\omega_1^2-V_{ml}\right)\varphi_1=0~,\\
\frac{\text{d}^2 \varphi_2}{\text{d}r_*^2}+\left(\omega_2^2-V_{ms}\right)\varphi_2=0~,
\end{eqnarray}
where $r_*=\int \frac{\text{d}r}{f(r)}$ is the tortoise coordinate and
\begin{eqnarray}
V_{ml}&=&-\frac{f(r)\left(f(r)-2rf'(r)\right)}{4r^2}~,\\
V_{ms}&=&f(r)m^2-\frac{f(r)\left(f(r)-2rf'(r)\right)}{4r^2}~,
\end{eqnarray}
are the effective potentials of the massless scalar field and the massive scalar field, respectively.

\begin{figure}[h]
\centering
 \subfigure[$V_{ml}(r)$]{
 \label{fig3a}\includegraphics[width=.45\textwidth]{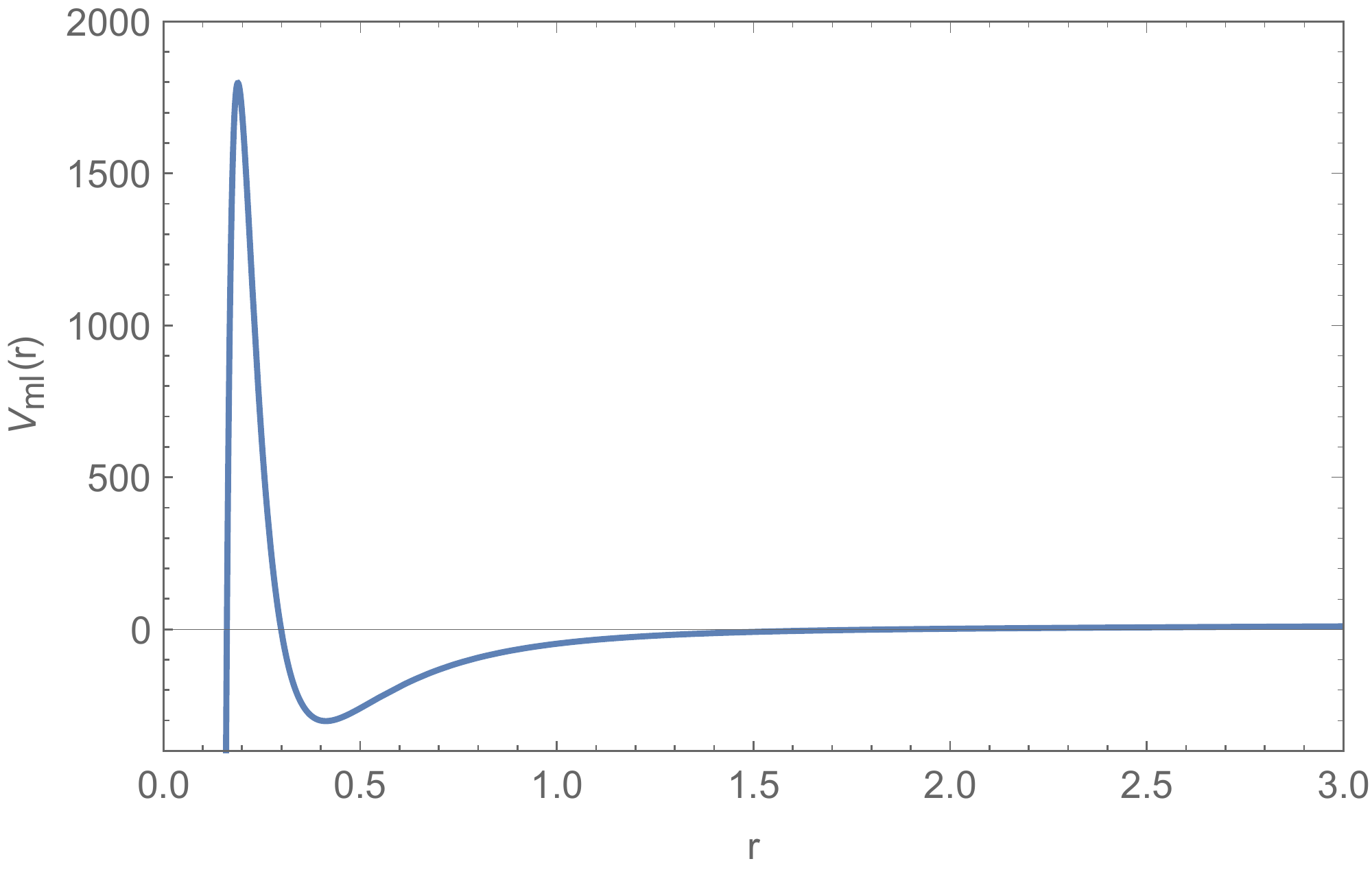} }
\subfigure[$V_{ms}$]{
 \label{fig3b}\includegraphics[width=.45\textwidth]{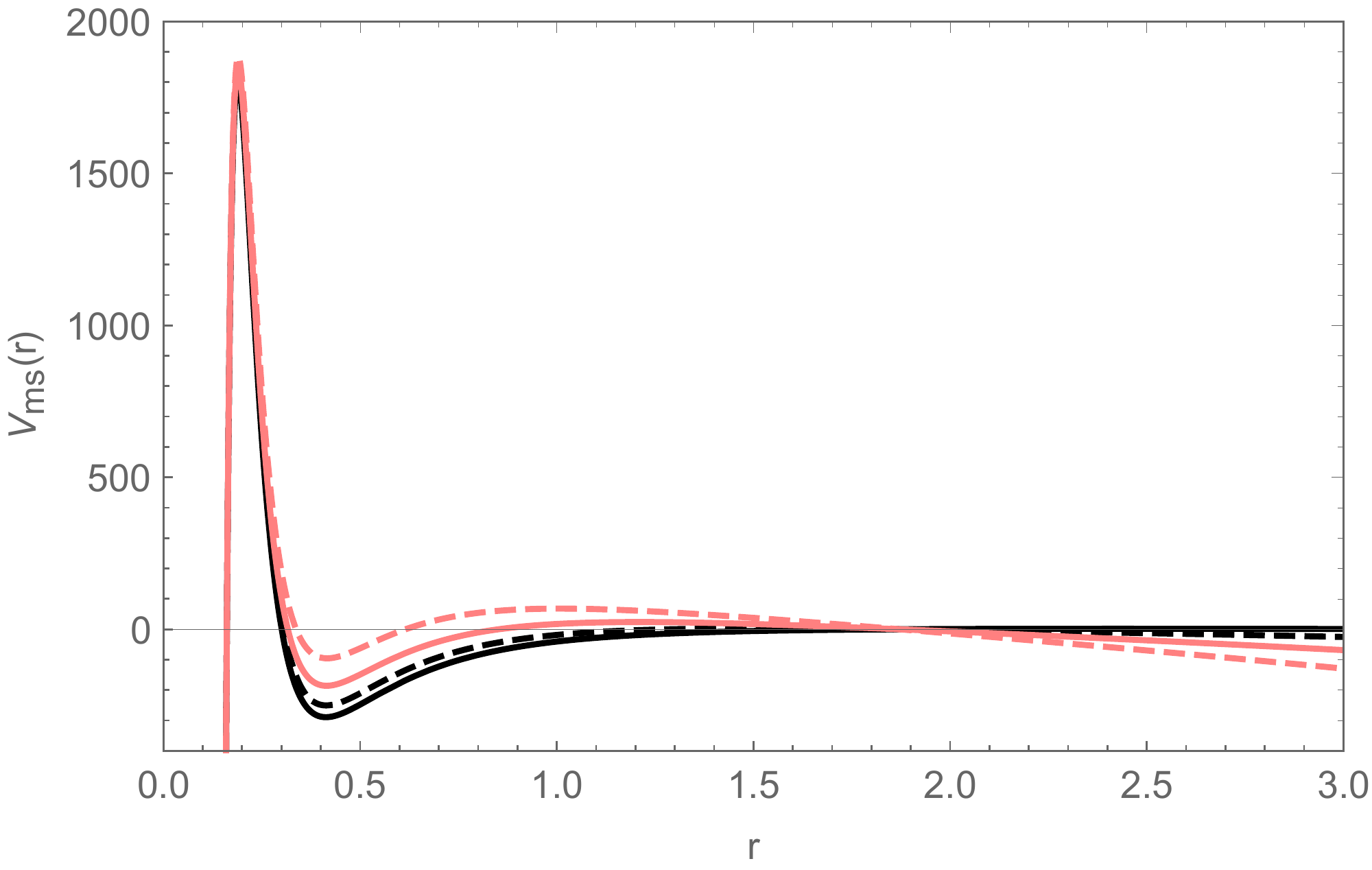} }
\caption{The effective potentials for (a) massless scalar field, (b) massive scalar field, with the black curve, black dashed curve, pink curve and  pink dashed curve represent scalar mass $m=1,2,3,4$ respectively. Here we take $\Lambda=1$, $B=-5$, $Q=2$ and $\beta=-1$. }
\label{fig:dSV}
\end{figure}

\begin{figure}[h]
\centering
\subfigure[$V_{ml}(r)$]{
 \label{fig4a}\includegraphics[width=.45\textwidth]{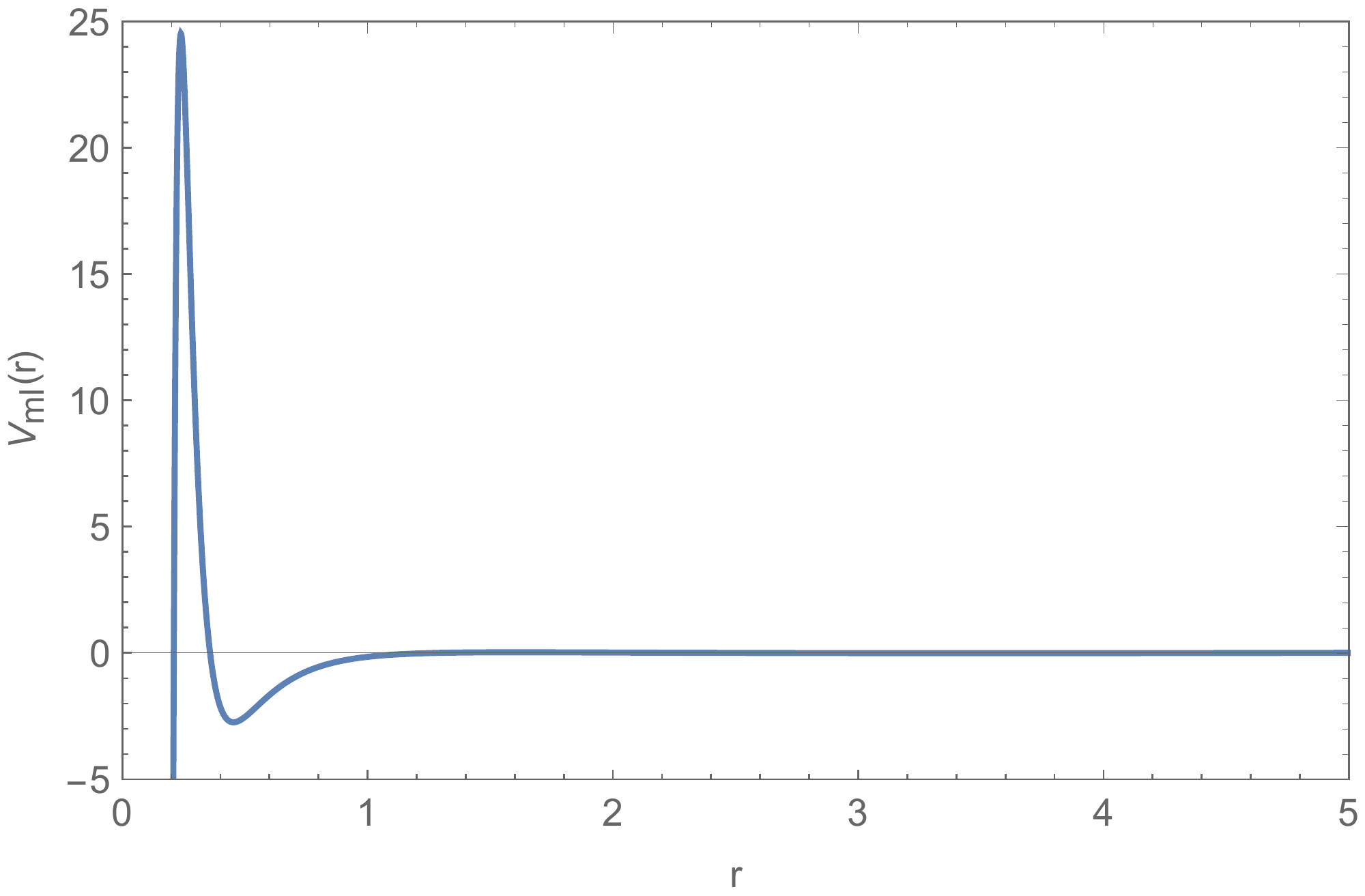} }
\subfigure[$V_{ms}(r)$]{
 \label{fig4b}\includegraphics[width=.45\textwidth]{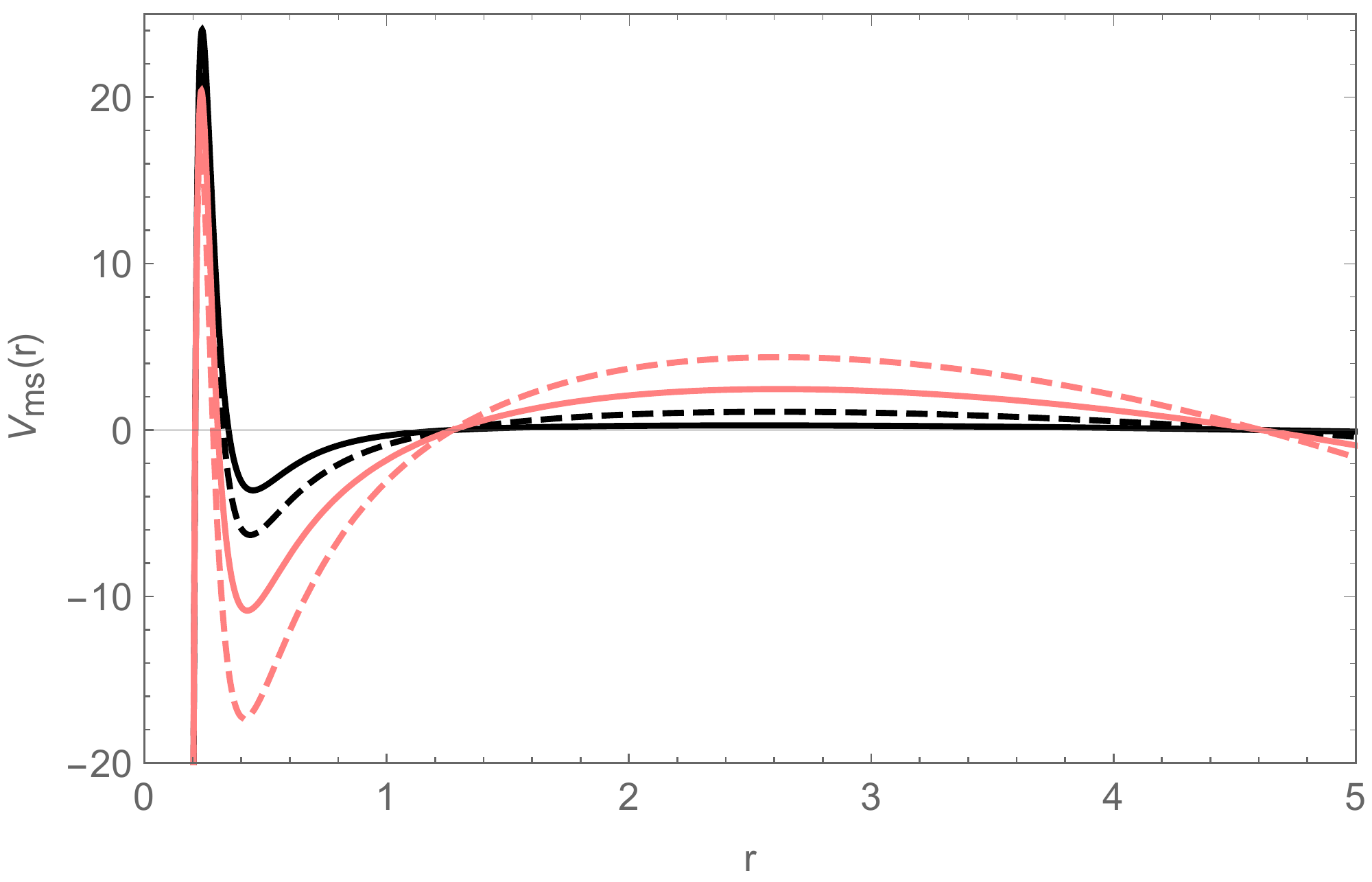} }

\caption{For comparison: the metric function of Reissner-Nordstr\"om-dS black hole is $f(r)=-\frac{2 M}{r}+\frac{Q^2}{r^2}-\frac{\Lambda  r^2}{3}+1$. Here we take $\Lambda=0.1$, $M=0.7$ and $Q=0.5$, which give Cauchy horizon $r=0.21$, event horizon $r=1.27$ and cosmological horizon $r=4.61$. Figure (a) corresponds to the effective potential of a massless scalar field, while figure (b) depicts effective potential for massive scalar fields: the black curve, black dashed curve, pink curve and  pink dashed curve represent scalar mass $m=1,2,3,4$ respectively. }
\label{fig:SdS}
\end{figure}

We plot these two effective potentials in Fig. \ref{fig:dSV}. From the figures we can see that there is a negative potential well outside the black hole event horizon. Therefore this dS black hole spacetime is maybe unstable under the massless scalar perturbations and massive scalar perturbations within some ranges of parameters. But this does not mean that the three-dimensional charged dS black hole we found is definitely unstable. This behavior is similar to that of four-dimensional Reissner-Nordstr\"om dS black hole under neutral scalar perturbation (see Fig. \ref{fig:SdS}), where the potential is  negative outside the black hole horizon.  Possible instabilities of quasinormal and superradiant spectrum usually appear in such regions in four-dimensions  with negative potential, see for example \cite{Zhu:2014sya}.

For comparison, we also plot the simplest BTZ and charged BTZ cases.

\begin{figure}[h]
\centering

\subfigure[$V_{ml}(r)$]{
 \label{fig5a}\includegraphics[width=.45\textwidth]{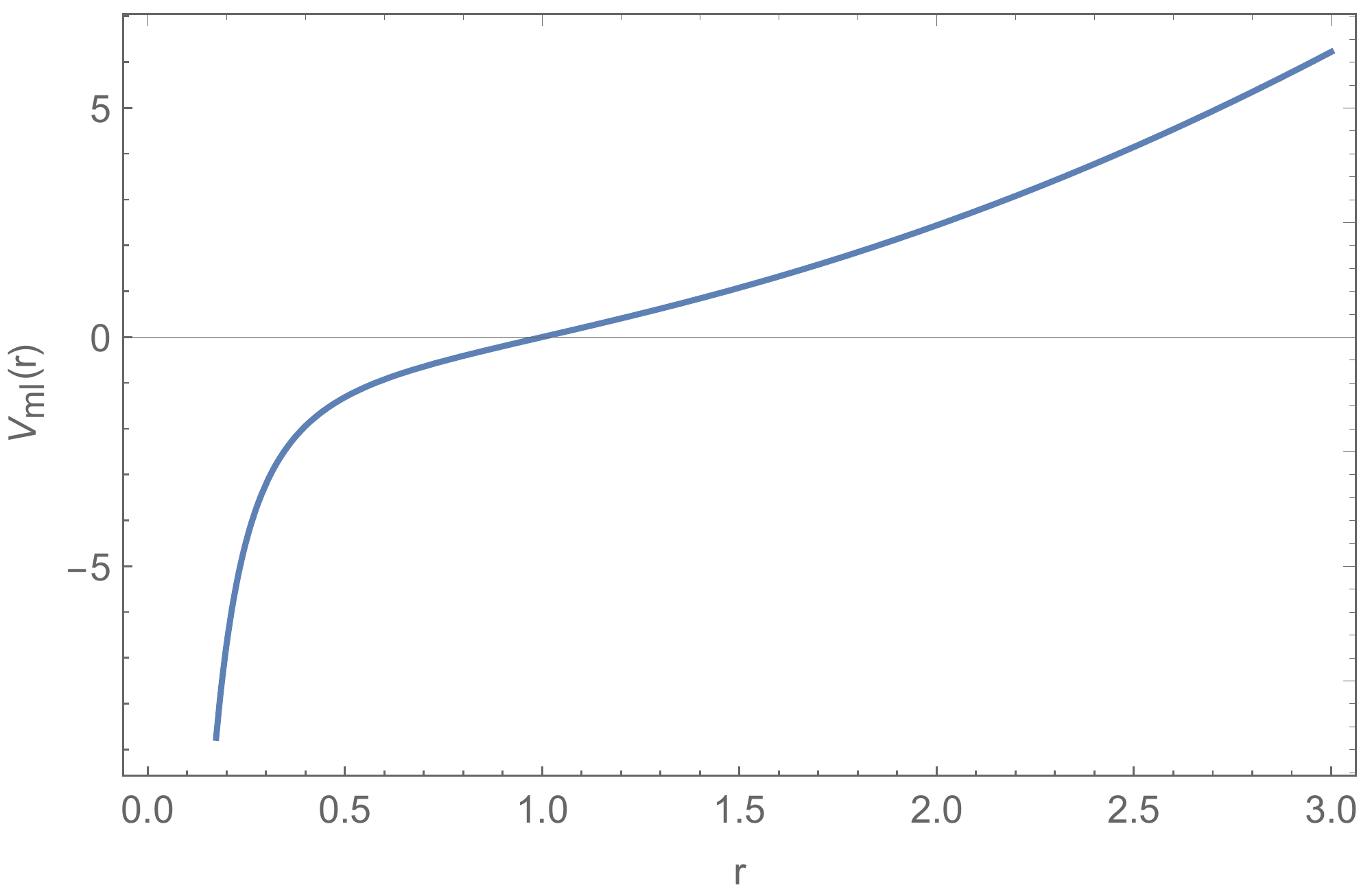} }
\subfigure[$V_{ms}(r)$]{
 \label{fig5b}\includegraphics[width=.45\textwidth]{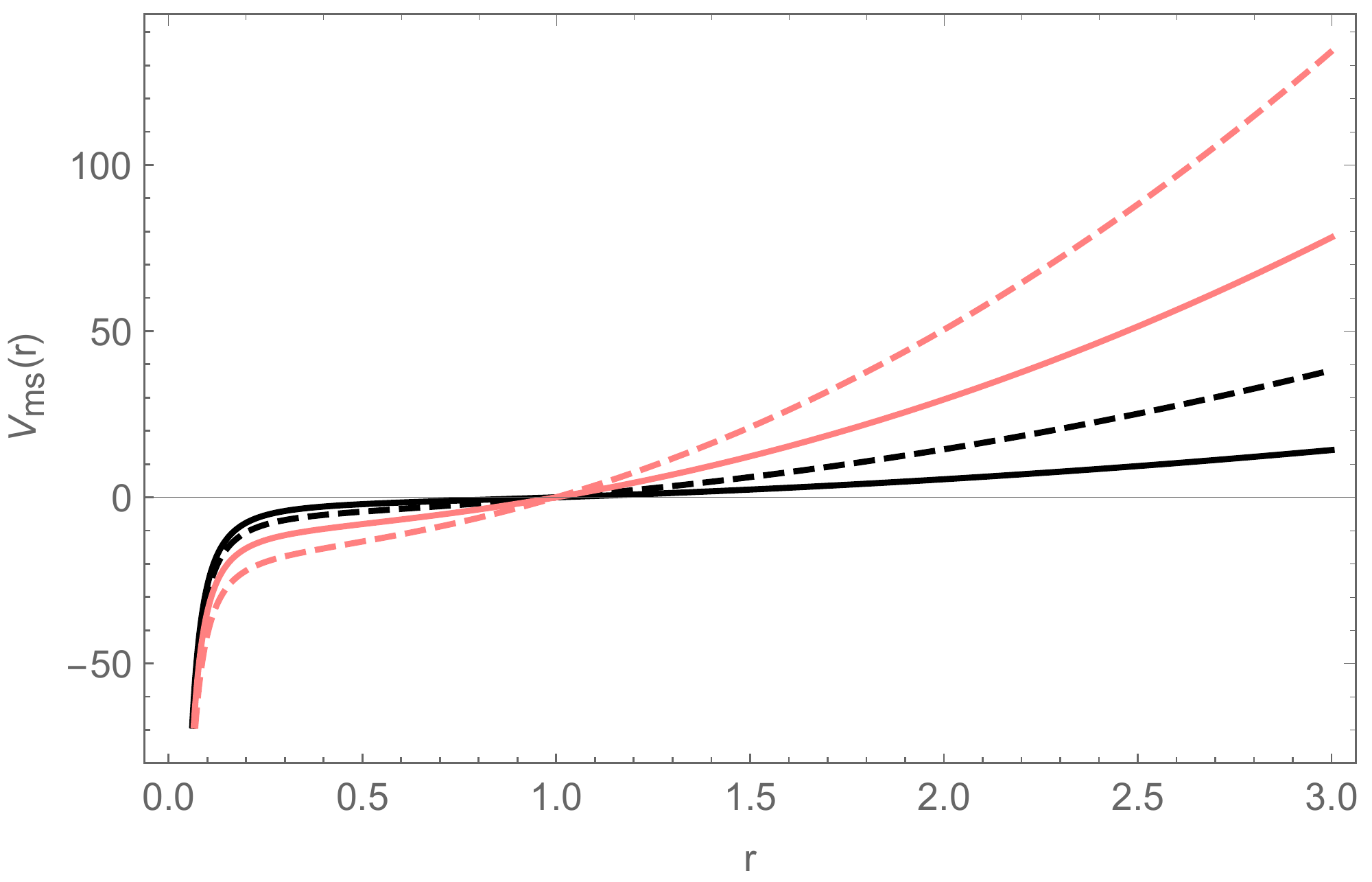} }
\caption{The metric function of BTZ black hole is $f(r)=\frac{r^2}{\ell^2}-M$. Here we take $\ell=1$ and $M=1$. In figure (a), we have the effective potential of a massless scalar field. In figure (b), we have the effective potential for massive scalar fields. The black curve, black dashed curve, pink curve and  pink dashed curve represent scalar mass $m=1,2,3,4$ respectively. }
\label{fig:BTZ}
\end{figure}

\newpage

\begin{figure}[h]
\centering
\subfigure[$V_{ml}(r)$]{
 \label{fig6a}\includegraphics[width=.45\textwidth]{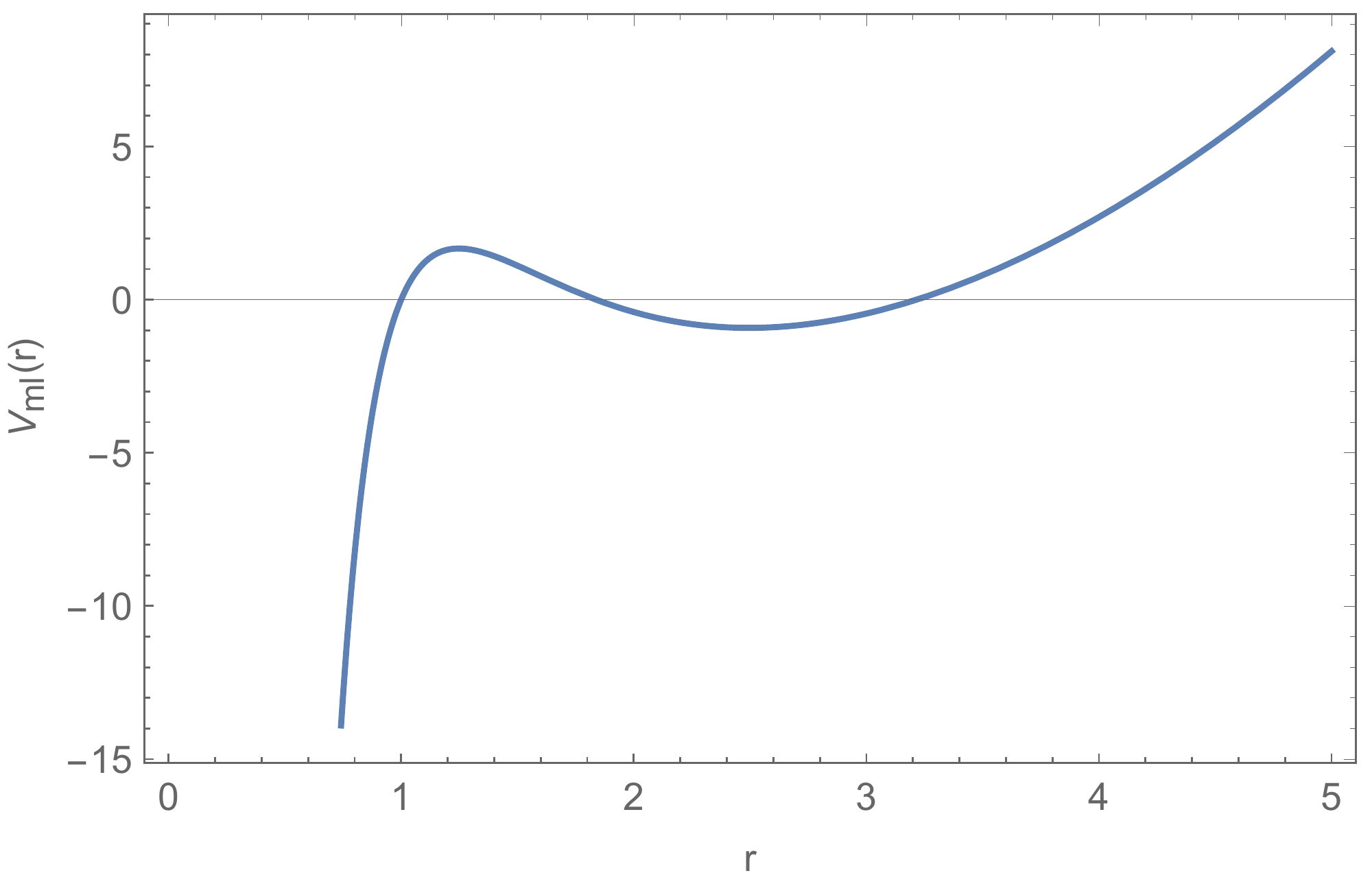} }
\subfigure[$V_{ms}(r)$]{
 \label{fig6b}\includegraphics[width=.45\textwidth]{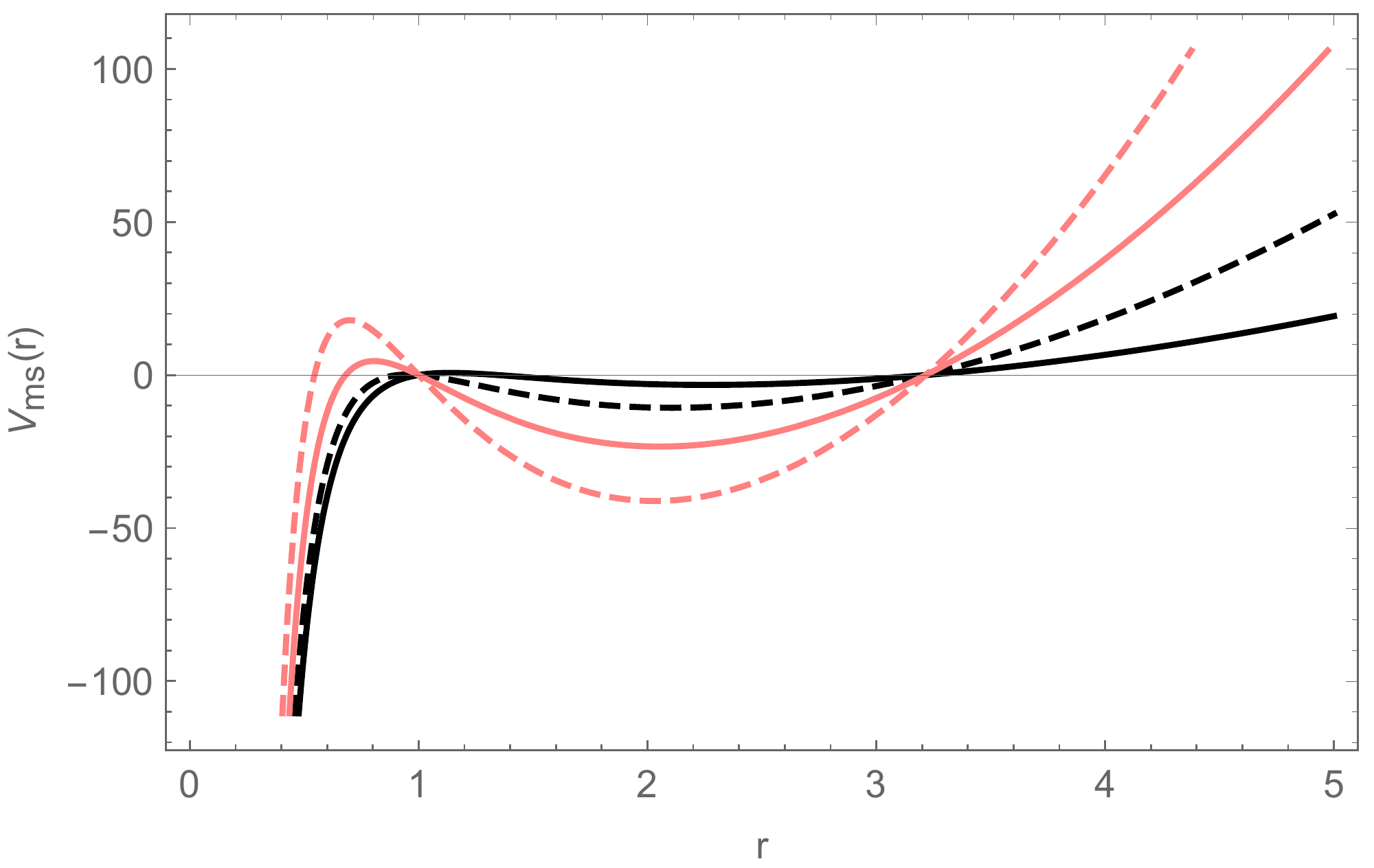} }

\caption{The metric function of charged BTZ black hole is $f(r)=\frac{r^2}{\ell^2}-M-2q^2 \ln{\frac{r}{\ell}}$. Here we take $\ell=1$, $M=1$ and $q=2$.
Figures (a) and (b) show, respectively, the effective potential for massless and massive scalar fields.
In  figure (b), the black curve, black dashed curve, pink curve and  pink dashed curve represent scalar mass $m=1,2,3,4$ respectively. }
\label{fig:CBTZ}
\end{figure}

From these plots we  see that the spacetime is more stable with larger mass of the scalar field in BTZ and charged BTZ backgrounds, while it becomes less stable with the increase of the mass in the (2+1)-charged dS black hole spacetime.



\newpage

\section{Conclusions}

The main purpose of our manuscript is to find a general black hole solution from the field equations when a scalar field is non-minimally coupled to Einstein's gravity in the presence of a cosmological constant. As it is well known that in  2+1 dimensional gravity only AdS black hole solutions are known. Our motivation was to make a systematic study of
black hole solutions in 2+1 dimensions with as few assumptions as possible. We found a solution of the Einstein equations and the Klein-Gordon equation with a general scalar potential for arbitrary coupling constants and integration parameters. This solution is new in the literature and based on the form of this potential we reproduced all the known black hole solutions in 2+1 dimensions and also we produced new solutions which were not found previously.

It was argued in \cite{Fan:2015tua} that it is very unlikely to find an exact solution for a random potential $V$. Fortunately we obtained the general potential with arbitrary coupling constant and four free integration constants, and based on this general potential we found exact solutions of the metric function. Starting from this general potential we reproduced all the known black hole solutions in 2+1 dimensions, besides we have also got some new solutions in our formalism. Considering that the integral in the expression of potential may bring divergent points, we also proved the continuity condition (\ref{continuity}) for the potential and metric function with arbitrary coupling constant.
We analyzed the asymptotic behaviors of potential and metric functions for different ranges of $\delta$. We found that for any range of values  of $\delta$  the asymptotic behavior of metric is $f(r\to\infty)=-C_1 r^2=-\Lambda r^2$.

Further, we discussed the influence of the coupling constant on the sizes of the event horizon and the form of the resulting scalar potential by plotting the figures of the metric and potential functions. Although not all integrals of potential $V$ can be integrated out explicitly, a list of exact analytic solutions can be obtained for $\delta=n, 1/n, -n, -1/n$ where $n=1,2,3,4...$. These solutions have a simple form of the scalar field but the metric function and the scalar potential are complicated functions with arbitrary parameters. However, an appropriate choice of parameters involved can bring these solutions to a simpler form. The properties of these black holes will be interesting to study in future works.

We also attempted in this general framework to find asymptotically de Sitter black hole solutions. We showed the requirement that the parameters of our theory should satisfy appropriate constraint conditions, which does not allow dS black hole solutions in (2+1)-dimensions.
Likewise, we found that there cannot exist asymptotically flat black holes.  As is well known in (2+1)-dimensions there are strong constrains coming from the energy conditions  \cite{Ida:2000jh} for the existence of black holes, so this is to be expected, though it is
still interesting to see how our construction fails explicitly. However, when an electromagnetic field is introduced, surprisingly it is possible to find a charged-dS black hole solution in some ranges of parameters. Nevertheless, this dS black hole solution is maybe unstable under scalar perturbations. It would be interesting to further explore the energy conditions, and see how such a black hole evades the previous no-go results.

\section{Acknowledgement}
This work is partially supported by the National Natural Science Foundation of China (NNSFC).
Y.C.O thanks NNSFC (grant No.11705162) and the Natural Science Foundation of Jiangsu Province (No.BK20170479) for funding support.
 E.P. acknowledges the hospitality of the School of Physics and Astronomy,
of the Shanghai Jiao Tong University where part of this work was carried out.
The authors also thank Zhong-Ying Fan, Dong-Chao Zheng and Bo-Liang Yu for useful discussions.

\end{document}